%% file: main.tex
\documentclass[acmsmall]{acmart}
\AtBeginDocument{%
  \providecommand\BibTeX{{%
    \normalfont B\kern-0.5em{\scshape i\kern-0.25em b}\kern-0.8em\TeX}}}

\setcopyright{acmcopyright}
\copyrightyear{2025}
\acmYear{2025}
\acmDOI{XXXXXXX.XXXXXXX}

\acmConference[CSCW 2025]{Make sure to enter the correct
  conference title from your rights confirmation emai}{}{}
\acmBooktitle{CSCW 2025} 
\acmPrice{15.00}
\acmISBN{}

\usepackage{multirow}
\usepackage{graphicx}
\usepackage{makecell}

\begin{document}

\title[Deaf-Centric Mixed-Reality Design]{Customizing Generated Signs and Voices of AI Avatars: Deaf-Centric Mixed-Reality Design for Deaf-Hearing Communication}

\author{Si Chen}
\email{sic3@illiois.edu}
\orcid{0000-0002-0640-6883}
\affiliation{%
  \institution{School of Information Sciences, University of Illinois Urbana-Champaign}
  \city{Champaign}
  \state{Illinois}
  \country{USA}
  \postcode{61802}
}

\author{Haocong Cheng}
\email{haocong2@illinois.edu}
\affiliation{
  \institution{School of Information Sciences, University of Illinois Urbana-Champaign}
  \city{Champaign}
  \state{Illinois}
  \country{USA}
}

\author{Suzy Su}
\email{xiaoyus4@illinois.edu}
\affiliation{
  \institution{School of Information Sciences, University of Illinois Urbana-Champaign}
  \city{Champaign}
  \state{Illinois}
  \country{USA}
}

\author{Stephanie Patterson}
\affiliation{
  \institution{University of Illinois Urbana-Champaign}
  \city{Champaign}
  \state{Illinois}
  \country{USA}
}

\author{Raja Kushalnagar}
\email{raja.kushalnagar@gallaudet.edu}
\affiliation{
  \institution{Gallaudet University}
  \city{Washington}
  \state{District of Columbia}
  \country{USA}
}

\author{Qi Wang}
\email{qi.wang@gallaudet.edu}
\affiliation{
  \institution{Gallaudet University}
  \city{Washington}
  \state{District of Columbia}
  \country{USA}
}

\author{Yun Huang}
\email{yunhuang@illinois.edu}
\affiliation{
  \institution{School of Information Sciences, University of Illinois Urbana-Champaign}
  \city{Champaign}
  \state{Illinois}
  \country{USA}
}

\begin{abstract}




This study investigates innovative interaction designs for communication and collaborative learning between learners of mixed hearing and signing abilities, leveraging advancements in mixed reality technologies like Apple Vision Pro and generative AI for animated avatars. Adopting a participatory design approach, we engaged 15 d/Deaf and hard of hearing (DHH) students to brainstorm ideas for an AI avatar with interpreting ability (sign language to English, voice to English) that would facilitate their face-to-face communication with hearing peers. Participants envisioned the AI avatars to address some issues with human interpreters, such as lack of availability, and provide affordable options to expensive personalized interpreting service.  Our findings indicate a range of preferences for integrating the AI avatars with actual human figures of both DHH and hearing communication partners. The participants highlighted the importance of having control over customizing the AI avatar, such as AI-generated signs, voices, facial expressions, and their synchronization for enhanced emotional display in communication. Based on our findings, we propose a suite of design recommendations that balance respecting sign language norms with adherence to hearing social norms. Our study offers insights on improving the authenticity of generative AI in scenarios involving specific, and sometimes unfamiliar, social norms.
\end{abstract}



\begin{teaserfigure}
  \includegraphics[width=\textwidth]{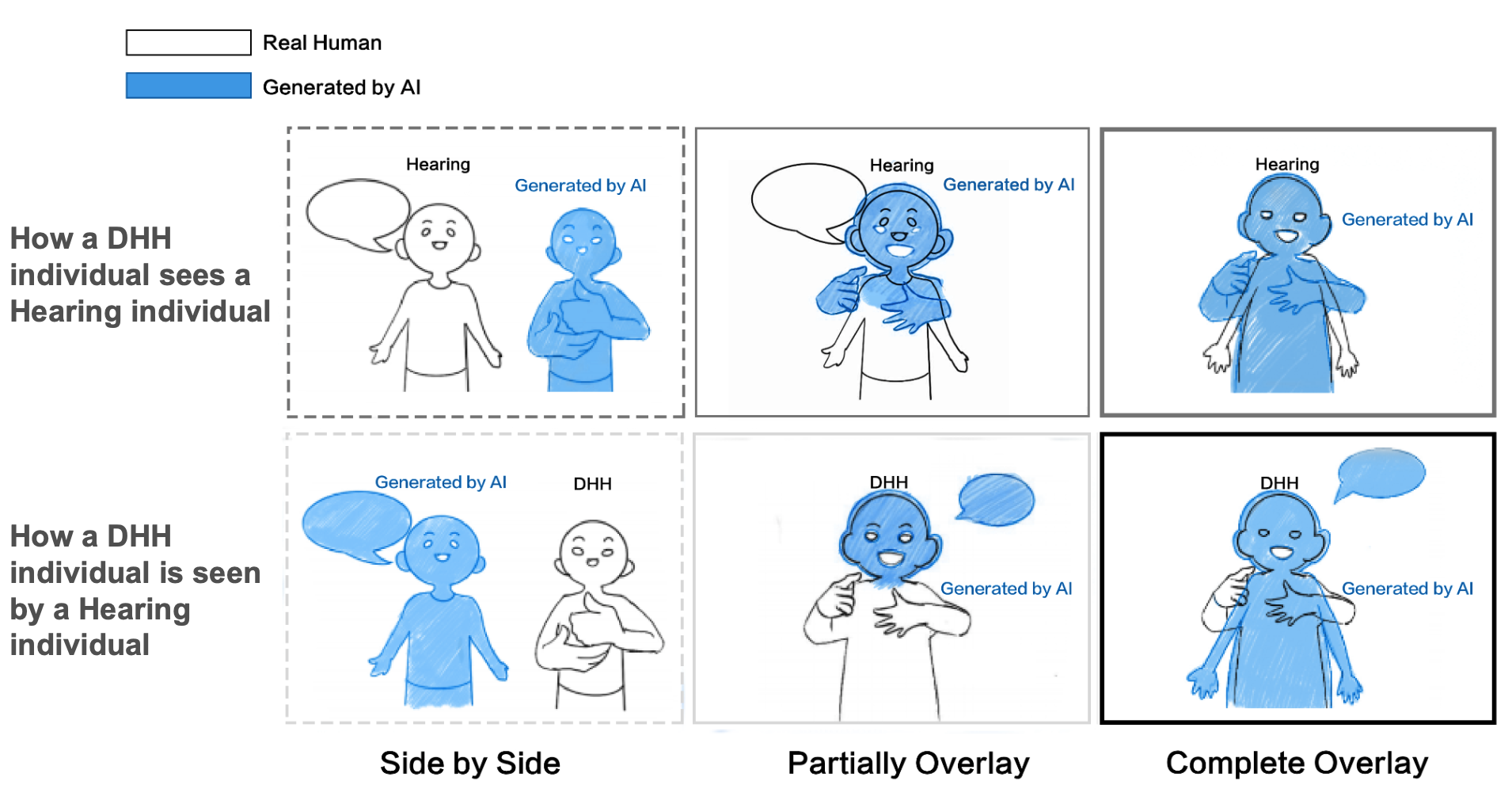}
  \caption{
 The illustration of mixed-reality (MR) designs that was explored in our study to support DHH-hearing communication. The top row illustrates scenarios that the DHH users would like to view on their MR glass, while the bottom row shows what they wish their hearing counterparts to see on theirs in the communication exchange. For both DHH and hearing individuals, there are three types of AI Avatar/filters that provides interpreting service , listed from left to right: 1) no overlay, which projects a separate avatar next to the actual person; 2) partial overlay, which covers a hearing individual's face and upper body for sign language when they speak and a DHH individual's face for spoken language when they sign; and 3) complete overlay, where a digital avatar entirely overlays the real person. The three overlay options illustrated different ways of blending for virtual objects and real-life objects in MR. The middle option, "partially overlay," explained the various ways the virtual and reality can be blended as the core of MR. Participants were informed they could use the overlay options as starting points to brainstorm additional features and control preferences, rather than simply picking a preferred one.
 Additionally, participants expressed a desire for features that allow them to regulate the AI-generated content to better comply with social norms. This figure was illustrated by the research team and presented to our participants during Session 1 of the design study to introduce and explain MR technology. 
 }
  \Description{xxx}
  \label{fig:teaser}
\end{teaserfigure}


\maketitle

\input{1-introduction}
\input{2-literature}

\input{3-method}

\input{6-findings}

\input{7-finding-2}

\input{8-discussion}
\input{9-conclusion_and_limitation}

\bibliographystyle{ACM-Reference-Format}

\end{document}

%% file: 1-introduction.tex
\section{Introduction}
In everyday life, many d/Deaf and Hard of Hearing (DHH) \footnote{In this paper, we use DHH to refer to the community of people with a wide range of hearing loss and cultural experiences. As is standard in the DHH community, we use deaf to refer to the medical or audiological condition of being deaf, and we use Deaf (capitalized) to refer to individuals who are involved in the Deaf community and use ASL \cite{pudanssmith2019}} individuals experience barriers in face-to-face communication in a predominantly hearing world. DHH individuals regularly employ visual, auditory, or tactile modes in face-to-face communication.  The most common visual communication modes include Sign Language  -- a distinct language in its own right with phonological, morphological, syntactic, etc., structure \cite{valli2000linguistics}. According to the World Federation of the Deaf \footnote{https://wfdeaf.org/}, there are more than 70 million deaf people worldwide, and, collectively, they use more than 300 different sign languages which are considered as fully fledged natural languages that are structurally distinct from the spoken languages. 
Our interaction design primarily focuses on this specific DHH population -- DHH individuals who sign.

American Sign Language (ASL) use in the US is substantial. 
More than 500,000 individuals use ASL as their primary mode of communication in face-to-face interactions \cite{articlele}. When DHH individuals use ASL to communicate with hearing people face-to-face, sign language interpreters are frequently used to accommodate the two-way communication process. Human interpreters primarily translate spoken language into sign language and/or vice versa. Their role is more than just translating but also facilitating effective and meaningful communication while upholding ethical standards (e.g., maintaining strict confidentiality about the information they encounter during their work) and respecting the unique needs of the individuals involved \cite{yabe2020healthcare, nicodemus2015}.  

However, human interpreters need to be pre-scheduled \cite{seita2020} and can be imperfectly trained or inexperienced \cite{nicodemus2015}, thus it is implausible to have enough high-quality interpreters accessible for every space and situation needed. Within the HCI community, there is a growing emphasis on the use of technology to support DHH-hearing communication. A long-studied technology is automatic speech recognition technology, which transcribes a hearing person's auditory speech to captions  \cite{berke2019} \cite{alonzo2020} \cite{seita2020}. 
Meanwhile, as stated before, DHH individuals who sign, prefer signed language over spoken language in face-to-face communication if supported, as it affords them a natural and intuitive way to express themselves in a conversation\cite{mack2020social}.

Given advancements in mixed reality (MR) technologies, such as Apple Vision Pro \footnote{\url{https://www.apple.com/apple-vision-pro/}}
, and generative AI for animated avatars \cite{10193198, syeda2023development, grewe2021statistical}
, this study explores novel interaction design to support communication between individuals with mixed abilities in face-to-face communication. 
We seek feedback and opinions from the DHH community regarding the integration of sign language-related technology with MR. As exemplified by Luo et al. in their work on signing avatars in classrooms \cite{luo2022avatar}, the avatars can be dynamically adjusted in terms of size and spatial positioning, catering to diverse scenarios and consequently, improving DHH students' classroom experience. In their study, the avatar interpreter enables one-way communication by translating the lecturer's voice message into sign language but student responses are ignored. In our design, we aim to provide two-way communication so both parties can converse and exchange their views seamlessly. The MR technology we envision is AI-based, designed for real-time ASL-English translation without prescheduling, addressing communication challenges with human interpreters. It might still be powered by human interpreters if available.    

Inspired by research through design fiction approaches (e.g.,\cite{designfiction2023,sondergaard2018intimate, pschetz2019autonomous}), we conducted participatory design with 15 DHH students to understand what an ideally inclusive world would look like using technology. Our aim was to provoke participants to envision AI to support DHH-hearing communication in MR. We also sought to identify any concerns our participants had about such technologies. Design fiction related forms of design research (e.g., speculative design, critical design,) are increasingly being used to investigate how emerging technologies, such as AI, shape human experience. These approaches also provide a focused context for debating their future application potential \cite{bleecker2022design, blythe2014research}. In this study, we address the following research questions : For DHH individuals who sign, 1) RQ1: With the envisioned MR technology with interpreting ability, what DHH-Hearing face-to-face communication challenges can be addressed? 2) RQ2: what MR overlay features are necessary to facilitate DHH-Hearing communication in MR? 3) RQ3: what design recommendations are derived for MR-supported DHH-Hearing communication?

Our design process is shown in Figure \ref{fig:processfig}. RQ1, answered in Session 1, focuses on the challenges in DHH-hearing communication that the MR technology can address. RQ2 is answered in Sessions 2 and 3, where participants propose features of MR overlay with interpreting ability to address certain challenges identified in RQ1. RQ3 is about other design recommendations concerning factors that impact the use of MR overlay with interpreting ability to support DHH-hearing communication. In Section 3, we discuss the recommendations made by our participants although they did not propose specific features in the study. Hence, we encourage researchers in this field to address and follow up on these recommendations in their future studies. 

Our work provides timely insights to the HCI community on supporting communication among individuals of mixed abilities. We identified preferences for integrating AI avatars with real-life images, from standalone AI interpreters to overlays with interpreting ability. Key findings highlight the importance of control features, such as tailoring signs, adjusting voices, and synchronizing facial expressions. Participants also offered design recommendations for MR technology to be accepted by both DHH and hearing users, emphasizing adaptability to diverse language preferences. This research advances inclusive and ethical human-AI interactions.

%% file: 2-literature.tex
\section{Related Works}

\subsection{Disability, Avatars, and Identity in Mixed-Reality (MR)}
The advancement of augmented reality (AR), virtual reality (VR), and mixed-reality (MR) technologies broadens research opportunities and the mode of interaction to encourage future avatar research. 
MR differs from VR and AR by merging the real and virtual worlds, allowing real-world objects to interact with virtual objects in real-time, thereby creating a seamless and interactive environment. Unlike VR, which immerses users in a completely virtual environment, and AR, which overlays virtual elements onto the real world, MR blends both realms to enable direct interaction between physical and digital objects
\cite{rokhsaritalemi2020review}. Furthermore, MR combines some of the benefits of AR and VR. Users, for example, do not have to purchase an expensive headset because they are optional, and can view MR created virtual elements that are perfectly embedded, i.e. cannot be separated, in the real object/environment \cite{mcmillan2017virtual}. 

Recent applications and designs of MR and AR to address accessibility problems have shown promise, including real-time captioning using the Vuzix glasses \cite{mathew2022aod}, as well as prototypes for sign language avatars that move around with the user's gaze \cite{luo2022} \cite{kercher2012}. These tools seek to alleviate the eye gaze strain that many individual experience when trying to take in all the visual information in a presentation or classroom setting, including the presenter or teacher, the presentation slides, and a personal notebook or computer. Especially when it is necessary to take in complicated and detailed visual diagrams, eye gaze strain presents a significant accessibility barrier for individuals \cite{behm2015}. Yang et al. proposed an avatar-based MR remote collaboration system that visualizes speech using spatial floating individual speech bubbles \cite{9090404}.

However, some risks to users' security and privacy exist when using MR. For example, because MR requires a large amount of information, including necessary and sensitive information, to provide service and allow users to customize settings, users may be unaware of this, but they may not want this latent information or contexts to be detected \cite{de2019security}. There are some methods for protecting users' privacy in MR applications including a variety of protections for data, input, output, user, and device, respectively. Under each category, users can select alternative systems and approaches to protect their privacy \cite{de2019security}. Participants in research on accessibility barriers for immersive technologies highlight issues with both software and hardware usability of AR/VR devices. Limited space for hearing aids in current bulky AR/VR systems is a major obstacle. Additionally, poor rendering of avatars and visual information, essential for lip and face movements, impedes DHH individuals' use of AR/VR technologies \cite{creed2023inclusive}. Other challenges include users having to sign with one arm or find a fixed location to sign with both hands \cite{berke2020chat}.

Furthermore, AR, VR, and MR technologies can now cover visible body parts ranging from hands and heads to the entire body, in a variety of styles such as cartoon and realistic \cite{10049669}. According to research, the user experience during AR/VR communication is usually better with generic realistic avatars \cite{reinhardt2020embedding} or agents \cite{huang2022proxemics} because eye contact is provided during the conversation, but realistic rendering styles are especially preferable for communication \cite{10049669}. As a result, users' digital self-identity has emerged as an important research topic. People in a social setting may need to become acquainted with one another and can interact with each other's avatars. Here's an example from Kundu et al. of how different identities, such as sexual orientation, can influence how people design their avatars or how users perceive them \cite{9974400}. Different cultures, genders, and emotional expressions will also influence individuals' decisions on whether to appear friendly in social settings\cite{huang2022proxemics}.

Users can communicate and interact with one another in digital worlds or applications using digital avatars \cite{nowak2018avatars}. Users can customize the appearance of their characters and use them on social media platforms such as iMessage and Instagram. According to Goffman, when people interact with one another, they want to control how they appear to others \cite{goffman1959presentation}. AR, MR, and VR devices and applications are becoming increasingly popular. More people with disabilities are using social VR now that it is available on the market \cite{zhang2022s}. These novel platforms provide excellent opportunities to benefit minorities, such as DHH users. Some researchers design and search for the best MR avatar configuration for DHH students, which includes a holographic sign language interpreter embedded in the MR space \cite{yang2022holographic}. Kaur and Kumar investigate and design \cite{kaur2016hamnosys, dhanjal2019comparative} a sign animation system for sign language automation that makes use of conversion between the Hamburg Notation System and the Signing Gesture Markup Languages.
Generative AI, as Gurung described, is designed to create new content based on existing data. Another example is LensaAI, which features a 'Magic Avatar' function that enables users to generate various styles of images from their uploaded photos \footnote{\url{https://prisma-ai.com/lensa}}. 
Mack et al. found participants with disabilities create and use avatars because using avatars can control how other people see them, deliver their variable needs and identities, and make communications easier since avatars are visual \cite{mack2023towards}.

\subsection{Current DHH-Hearing Communication Approaches}
Currently, the majority of accessible technology research is focused on improving the availability and quality of captions to provide individuals with access to auditory information. Studies often explore the impacts and considerations of ASR and captioning for DHH viewers \cite{mcdonnell2022, li2022, berke2019, seita2020}. Though ASR has greatly improved access to speech info for DHH users, it is important to note that captions are not the only or the most ideal options for two-way communication between a DHH and a hearing person. With captions, the hearing person may convey information to the DHH counterpart, but it does not work the other way around. 
With the advancements in VR, AR, and MR technologies, there have been significant innovations aimed at supporting DHH users, particularly focusing on speech and sound access. Guo et al. developed a HoloLens-based AR prototype that classifies and visualizes sound identity and location, in addition to providing speech transcription \cite{guo2020holosound}. Another work also focused on sound awareness in group settings \cite{jain2015head}. Jain et al. designed and evaluated a novel taxonomy for VR sounds, contributing to sound awareness in VR for DHH users \cite{jain2021taxonomy}.

Sign language is an important communication medium for DHH people, as captioning and sound awareness can be inaccessible or insufficient. In the US, DHH individuals born to deaf parents could learn ASL fluently, while those born to hearing parents often pick it up later in life and are capable in both ASL and spoken English \citep{mitchell2004parents, lu2016impact}. Most hard-of-hearing children initially learn a spoken language but might transition to ASL later \citep{mitchell2005parental}. In short, a large percentage of those individuals, whether acquired ASL early or late in life, gravitate to signing when they communicate with each other.  For this reason, many DHH individuals prefer using sign language interpreters when communicating with hearing people. To benefit DHH users who sign, there are VR studies on how sign language interpreters should be rendered (fixed-position or always-visible) and guided in VR \cite{anderton2022investigating}. It is important to note that Anderton's work displays sign language interpreters as videos, not generated 3D avatars. Additionally, we focus on interpreting as a two-way process, not just on displaying the human interpreter.

Sign language interpreters provide accessibility and a bridge between DHH and hearing individuals in face-to-face communication. However, this solution is not optimal as human interpreters can make mistakes and are not always available when needed. For example, in a study by Nicodemus and Emmorey, novice ASL human interpreters tended to overrate their abilities in ASL, but their accuracy was inadequate when interpreting into ASL than into English \cite{nicodemus2015}. This suggests that newer ASL interpreters tend to make more mistakes when interpreting into ASL, which may impact a DHH person's ability to understand the interpreted message. Additionally, in a 2020 study analyzing complaints about human interpreters in healthcare settings, almost half of the complaints were about the lack of ASL interpreters availability when the service was called for, and a third complaints were about the poor quality or qualification of the interpreters who provided the service \cite{schniedewind2020}. This shows that the availability and quality of interpreters are not uniform which can lead to frustration and lack of accessibility for DHH individuals. There are also a limited number of interpreters (about 1 interpreter per 50 DHH people, per Registry of Interpreters for the Deaf, so for face-to-face meetings, there are typically not many interpreters close by, which means they are not immediately available on call. So it can be difficult to provide on-call interpreting. Because of availability, communication barriers and misunderstandings do occur when using ASL interpreters, and they are not a perfect solution for facilitating reliable face-to-face communication between hearing and DHH individuals.


We reviewed recent advancements encompassing sign language recognition, generation, translation, and animation. While these advancements hold immense promise, there is much work to be done before these technologies become more useful for providing accessibility. Recent work has improved Sign Language Production models by integrating a Frame Selection Network that enables precise sign alignment, as well as SIGNGAN for generating photo-realistic sign language videos \cite{saunders2022signing}. While this study achieved commendable advances in refining co-articulation and overall signing quality, it's crucial to acknowledge the constraints inherent in large-scale text-to-gloss translation, prompting the need for further progress in model architectures and datasets. A recent overview of the state of the Sign Language Translation field emphasized the importance of integrating sign language into existing machine translation technology while involving DHH end-users in the development of an SLT application \cite{decoster2023}. As this technology continues to develop, it is essential that members of the DHH community are involved in designing, testing, and applying this technology to ensure their culture and community is respected.

%% file: 3-method.tex
\section{Methods}
We used a participatory design approach to explore the accessibility challenges that DHH individuals face in daily life, and to design solutions using some real as well as fictional technological advancements. Participatory design is a collaborative approach that involves end-users in the design process \cite{spinuzzi2005methodology,schuler1993participatory}. Our study aims to create products and services that better meet the needs and expectations of target users by involving them in the design process and utilizing their respective knowledge and experiences. This is a university IRB-approved user study.

\textbf{Participant and Recruitment}.
We recruited DHH participants through email lists, social media posts, and word-of-mouth as a snowball sampling method.  
A total of 15 DHH participants, aged between 21 and 47, took part in the study. Five participants were identified as male, nine as female, and one as transmasc non-binary. Eleven participants were Deaf and four were hard of hearing (HoH). All participants in our study can sign, but signing was not necessarily their primary or only language used in everyday life. As for racial identity, eight were identified as White, while three were Asian, one Black, one Middle Eastern, one Latino, and one Mixed.
All participants were undergraduate or graduate college students who expressed a strong interest in VR/AR technology during our preliminary email exchange screening process. All of them were able to describe at least some experiences with AR/VR, such as a headset or a mobile phone. None of them, however, had design or development experience with AR/VR technology.

\subsection{Study Procedure}
\begin{figure*}[ht]
    \centering  
    \includegraphics[width=0.99\textwidth]{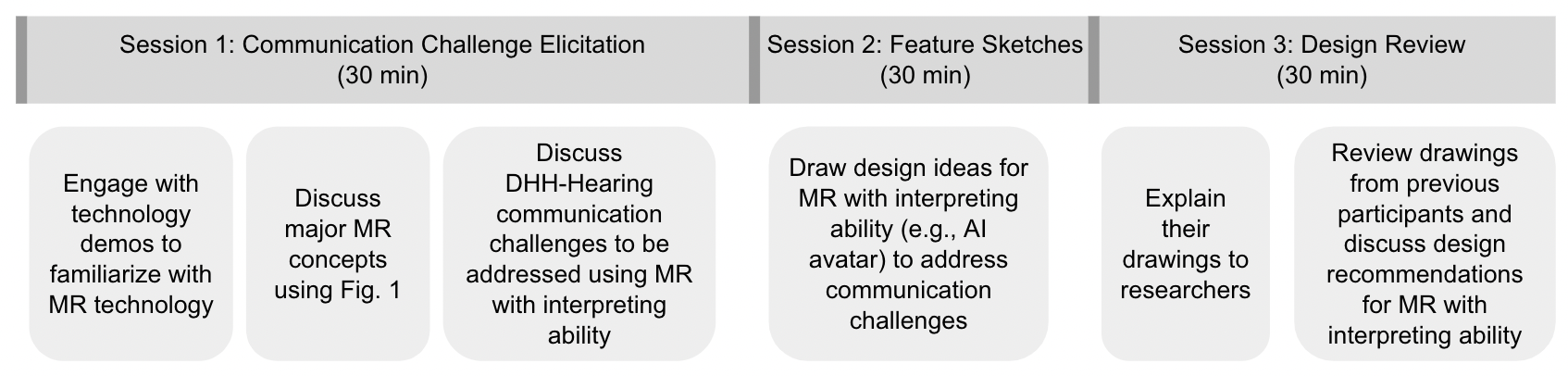} 
    \caption{An overview of the study procedure.}
    \label{fig:processfig}
    \Description{A process diagram showing the study procedure of the participatory design study. On the top are the titles of the three sessions. Session 1 is Communication Challenge Elicitation (30 minutes), with three blocks representing three activities: Engage with technology demos to familiarize with MR technology; Discuss major MR concepts using Fig. 1; and Discuss DHH-Hearing communication challenges to be addressed using MR with interpreting ability. Session 2 is Feature Sketch (30 minutes), with one block representing one activity: Draw design ideas for MR with interpreting ability (e.g., AI avatar) to address communication challenges. Session 3 is Design Review (30 minutes), with two blocks representing two activities: Explain their drawings to researchers, and Review drawings from previous participants, and discuss design recommendations for MR with interpreting ability.
}
\end{figure*}

In our three-session study, participants were introduced to relevant technology demos, asked to draw their ideal set-up for a one-on-one communication with a hearing person through sketches, and discussed their drawings with the researchers for an interview. The three sessions 1, 2, and 3 occurred sequentially. 
Afterward, participants were asked to complete a survey on their demographic information and language preferences.  
The study procedure is illustrated in Fig. \ref{fig:processfig}. Prior to each user study, we communicated with our participants via emails explaining the tasks to be completed during the study and presenting several questions to elicit their thoughts and interests.  We also told participants they would be asked to draw their ideas with any tools, digital or non-digital, of their own choice. 

Each user study was conducted individually, either in-person (seven participants) or over Zoom (eight participants) depending on their and the researchers' locations and schedules. Participants were compensated for their time (\$25 per hour).  The studies were conducted either through ASL (without an interpreter) or with ASL interpreters based on the availability of the research team members. Each study took about two hours, with a ten-minute break between sessions and whenever participants requested it. Two participants did not complete the study procedure within two hours due to fatigue, and they decided to rejoin the study at a later time to complete the remaining sessions.

\textit{Practice Study:} To refine the detailed session design, the research team practiced the design sessions with three participants. Initially, researchers found that even DHH participants familiar with AR/VR had challenges envisioning MR technology. The concept of how virtual objects can overlap with real-life objects was hard to envision visually. Additionally, none of the participants had seen AR/VR apps for supporting DHH-hearing communication, and they reported limited trust in ASL to English interpreting via AI. These led us to create Figure \ref{fig:teaser} (to be elaborated in section 3.1.1) to support the discussion and help explain MR overlay concepts visually. We conducted a second round of practice with two of the three participants to gather feedback on the revised study procedure. These participants had similar backgrounds to the formal study participants.

\subsubsection{Session 1: Communication Challenge Elicitation} \label{demo}

To familiarize the participants with both existing and fictional technological advancements that could enhance their communication accessibility, they were introduced to various technologies related to ASL-English interpreting and AR avatars. First, participants used the DeepMotion \footnote{\url{https://www.deepmotion.com/}} website to create an avatar with a realistic face and then uploaded an ASL video of themselves to be animated using the DeepMotion avatar they created. 
Next, participants viewed the Unreal MetaHuman \footnote{\url{https://www.unrealengine.com/en-US/}}, which portrayed the hyperrealistic facial expression animations that were now available. Finally, we showed a video of the Apple Vision Pro \footnote{\url{https://www.apple.com/apple-vision-pro/}} to participants and encouraged them to imagine using wearable technology to enhance accessibility. Participants were also given the option to view an example of sign language translation using the technology from \cite{kosa2023asl}, which leveraged existing machine learning methods that developed a mechanism to navigate and sign a consent form using ASL. 


As informed by our practice studies, we then asked each participant to recall a scenario from their daily life when they had good and bad communication experiences with a hearing person and when an ASL-English interpreting service could be beneficial. Most participants described situations where a virtual interpreter or interpreting service in the form of a 3D avatar would be beneficial. Then, to present a visual demonstration of the MR technology, the researchers proceeded by presenting Figure \ref{fig:teaser} to the participants and explained the three available overlay options with two views (DHH individual seeing and being seen by hearing individual). The three available overlay options showed different approaches of blending virtual objects in real-world physical spaces for an immersive interaction experience, which is the core concept of MR technology. The ``side by side'' and ``complete overlay'' options demonstrated two possible placements of the interpreting service avatars, whereas the ``partially overlay'' option was used to inspire the participants that the avatar might use a mid-ground design beyond the other two options.
Participants were asked to express their preferences of these options and the reasons behind their choices. 
Participants were also made clear that they were not confined to the options presented in Figure \ref{fig:teaser}; instead, they should consider using them as starting points to brainstorm additional features and related controls if necessary, as we proceeded to Session 2. 

\subsubsection{Session 2: Feature Sketches}  After identifying the scenario(s) and being familiarized with the technologies, participants were asked to create visual representations of how the technology could be integrated into the scenario(s) to improve communication with hearing individuals. We required them to think from both the DHH's view and the hearing's view for two-way communication. We also encouraged them, though not required, to think from a third-person point of view on how both DHH and hearing individuals could be using the technology to communicate. 

During the sketching process, we prompted participants to think extensively and to ignore technological constraints (assuming the glass device is affordable and widely adopted by the general public). We emphasized that while the study primarily centered on face-to-face conversations involving one DHH individual with one hearing individual (though in the future, there might be a variety of possibilities with technology advancement and affordability), participants had the option to explore group discussions in public spaces (involving more than two people) if they wished.

\subsubsection{Session 3: Design Review} 
To better understand participants' design preferences and any concerns they had about this type of technology, we asked participants to explain and clarify their drawings. We also showed them design examples (only drawings, but not explanations) from three anonymized previous participants to elicit their insights.The number of previous participants' designs was informed by the estimated session time in our practice studies. Note that five participants only reflected on two designs due to limited remaining time in this session. This activity encouraged participants to critically reflect on their specific challenges and contemplate the potential role of technology in mitigating these obstacles. 

\begin{figure}[!h]
    \includegraphics[width=.9\textwidth]{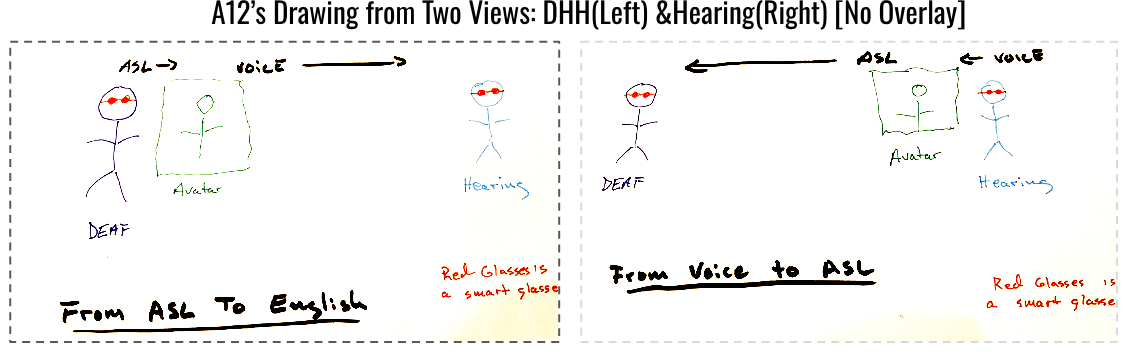}
    \caption{Example Feature Sketches from Two Perspectives: DHH Individual See Hearing and How DHH Individual Perceive Hearing See Them. 
    }
    \label{fig:A12}
    \Description{}
\end{figure}

An example of this critical thinking process is demonstrated by A12, whose drawing is shown in Figure \ref{fig:A12} (drawing on a piece of paper). From the DHH individual's view, he wants his ASL be translated to Voice (left), and from the hearing individual's view, he wants the other person's Voice be translated to ASL (right). His drawing shows he initially prefers a `no overlay' view in Figure \ref{fig:teaser} that looks more \textit{``equal''}. Then via design review with the researcher, he explained more nuanced thinking that `no overlay' might not be optimal than the overlay option because \textit{`` history consistently shows that hearing people who have never met a deaf person before always intended to look at them (interpreters) instead of making eye contact with the deaf person''}. 

\subsubsection{Additional Details for Study Setup}
While we kept the procedure and instructions the same for all participants via Zoom or in person, we also made accommodations to each participant's preferences. Zoom participants' studies were on average around 20 minutes longer than in-person participants' studies, as the tasks involving operating the technology and switching between different screens took longer. For in-person studies, the visual attention of researchers, interpreters, and participants was more easily guided, allowing everyone to have a clearer and quicker understanding of where and what to look at and how to communicate ideas. 

The major difference between Zoom and in-person sessions was related to the researchers' ability to fully observe how the drawing was done step by step in session 2. For Zoom participants, some were drawing outside of the camera’s view. For both in-person and Zoom participants, researchers did not ask for clarification during the drawing process because switching between drawing and signing was tiring for our participants. For Session 2, most Zoom participants drew on paper and took pictures to share their ideas via screen share. They then further annotated their drawings in front of the screen. Some participants also used Google Slides for easy edits and text annotations. Three participants drew on paper and showed it in front of the camera. All in-person participants drew on paper and then made further edits directly on their drawings in Session 3.

\subsection{Data Analysis} 
The collected data, including interview notes recordings, drawings, and field observation records, were analyzed using qualitative analysis. We conducted data analysis using reflexive thematic analysis, following the guidelines set forth by Braun and Clark \cite{clarke2021thematic}. Our approach to the data was grounded in a semantic and critical realist perspective. We employed an inductive method for all RQs. The interview transcripts from Session 1 were analyzed to identify the existing challenges in DHH-hearing communication that could be addressed by MR technology (RQ1). The drawings, interview transcripts and notes, and field observation records from Session 2 and 3 were analyzed together to understand the technical features to facilitate DHH-hearing communication through MR technology for DHH individuals who sign for communication (RQ2). Session 3 data was analyzed to understand as well as other design recommendations beyond technical features that should be considered in DHH-hearing communication through MR technology (RQ3).

For user studies that were conducted with no ASL interpreters, the interview dialog was transcribed to English by a HoH researcher. For user studies that were conducted with the interpreters, the user studies were automatically captioned in English scripts. The research team directly discussed and analyzed the system-captured caption scripts. The accuracy of the transcripts was verified by several researchers to ensure authenticity when some direct quotes were selected for inclusion in the paper. The entire research team was involved in the discussion and analysis of the transcripts. See Section \ref{sec:positionality} for the positionality of the research team.


The initial data analysis was spearheaded by the first author, who began by thoroughly reviewing interview transcripts, field observation notes, and survey data, documenting recurring patterns, and consolidating these observations into an initial codebook. This codebook was continuously refined with additional data from subsequent user study sessions and was reviewed and validated by other researchers. The final code was then applied to the data obtained from all study participants. At the team meetings, the researchers reflected on study session activities and discussed preliminary findings based on the artifacts from interviews, observation data, and summary information compiled by the first author to determine if we had effectively addressed the research questions. Through iterative discussions involving all coauthors, including several DHH faculty members experienced in human-computer interaction research, the first author grouped codes and data into the themes that now constitute the findings subsections of the study.

\subsection{Positionality} \label{sec:positionality}

The research team included members with diverse hearing abilities and sign language backgrounds. The first author of the paper was a hearing researcher and an ASL beginner. She led half of the studies and communicated with DHH participants via ASL interpreters. The other half of our user studies were led by a researcher who is HoH and fluent in ASL, and she communicated with DHH participants directly using ASL (with no interpreter). Another hearing researcher who is intermediate in ASL, a Deaf professor, and a hearing professor fluent in ASL attended the research sessions as observers and kept field observation notes.  Three additional hearing researchers participated in the data analysis process. 


%% file: 6-findings.tex
\section{Findings}
We report findings in the below sections organized based on research questions and their underscore themes derived from extensive data analysis from multiple sources to ensure data triangulation. For DHH individuals who sign, we address the following three RQs.  

\subsection{RQ1-With the envisioned MR technology that can interpret, what DHH-Hearing face-to-face communication challenges can be addressed?} \label{sec:rq1}
In this study, all DHH participants expressed enthusiasm about the envisioned technology which seemed to hold promise to allow both DHH and hearing individuals to conduct face-to-face communication in their natural language - ASL for DHH individuals who sign and spoken English for hearing individuals as illustrated in Figure \ref{fig:A12}. The proposed technology will interpret ASL and English, and then generate an AI signing overlay and AI voicing overlay simultaneously for the communicating pairs to send and receive messages in their own preferred languages. Below describe the communication challenges in DHH-Hearing face-to-face interactions that can be addressed by the envisioned technology.

  \begin{figure}[!h]
    \includegraphics[width=0.8\textwidth]{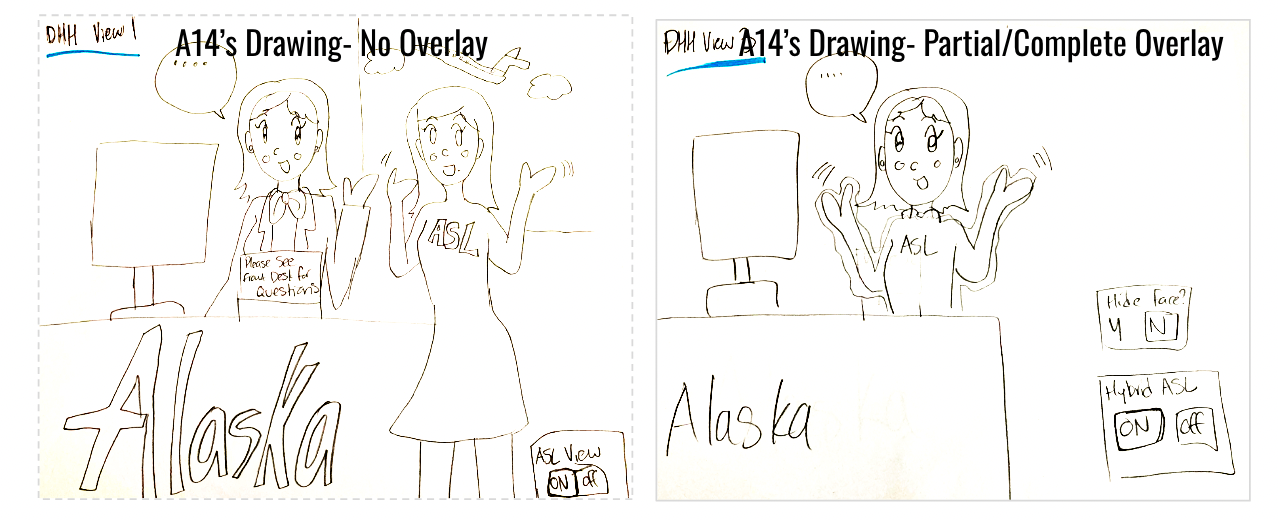}
    \caption{A14's Feature Sketches showing herself at the airport where she needs instant, fast, and accurate communication envisioned using proposed interpreting AI. She, as well as the majority of our participants, prefers `No Overlay' and `Partial Overlay' over `Complete Overlay.' In her illustration, she emphasizes the significance of DHH individual's control over whether to utilize the filter and fine-tune its settings, particularly in relation to deciding whether to apply a facial overlay onto the hearing individual in communication.}
    \label{fig:A14}
    \Description{}
  \end{figure}

\subsubsection{Extra Effort Entailed in Guesswork and Clarification} Our participants envisioned the proposed technology could provide a portable and readily available "interpreting service" at one's fingertips, which could save their effort in guessing or repeating for clarification in communication exchange with hearing people. Without a human interpreter, most DHH people relied on writing, lipreading, and voicing with \textit{``deaf accent''} when interacting with hearing people. Frequently, such exchange caused miscommunication and delayed realization of errors, \textit({``oh we're not talking about the same thing...hmmm...what happened...''}) and required more time to clarify. Several participants reported that they frequently experienced situations when hearing individuals were \textit{``inattentive to communication breakdowns''} and/or lacked patience for clarification, which left the deaf person to guess the message. A1 recounted a slow and \textit{``embarrassing''} ticket purchase experience at a train station with a hearing agent. Because the agent was impatient with him, repeated misunderstandings occurred, and the purchase took much longer to complete. This caused him additional anxiety as the people in the queue were getting upset about the delay. Another participant, A14, encountered a similar situation with an airport agent as depicted in Figure \ref{fig:A14}. A9 stated that sometimes she was unwilling to ask for clarity when she perceived that requesting clarification could disrupt (potential) relationship dynamics or slow the communication flow, so she just relied on guessing:

\begin{quote}
\textit{``It (communication with hearing individuals) is almost like `Mad Libs,' where I try to fill in the gap and put up with ambiguity. I read lips so sometimes that helps to understand the context better, even with a human interpreter already there...I was raised as a hearing child and not allowed to sign at home, I am used to doing that or I had to do that but that doesn't mean I like to do that...''} (A9)
\end{quote}
\subsubsection{Unqualified or Uncertified Human Interpreters} According to our participants, the proposed technology could provide a more reliable interpreting service than some human interpreters whose qualification or certification was not attested. It is known in the field that many interpreting service providers are having difficulties locating qualified or certified ASL interpreters and would send interpreters whose skills are not attested \footnote{\url{https://rid.org/certification/}}. A12 faced a medical situation where he decided to try an interpreter despite his proficiency in written English. However, the interpreter provided could only manage basic ASL, causing misunderstandings and disrupting communication with the doctor. Eventually, A12 had to ask the interpreter to stop and switch back to using written English. This experience shows that some places have a casual and dismissive attitude towards providing quality interpreting service, so the main burden of communication task once again fell on the DHH people in the dialog exchange: 

\begin{quote}
\textit{``The interpreter was not certified. And I was like, `wow.'... A lot of companies think that you can just call anyone and have them come in...I had to use a lot of time to write back and forth with the doctor. " }
(A12)
\end{quote}

\subsubsection{Lack of Affordable Personalized Experience w/ Human Interpreter} 
Another potential benefit of an automated interpreter highlighted by our participants was it could offer \textit{personalized} interpreting experience, which not every certified interpreter has the \textit{"experience and skillsets"} to achieve. Similar to how various regions exhibit distinct accents and dialects in spoken English, ASL also showcases individual or regional characteristics in its expression. While ASL is indeed a standardized language, different individuals have their unique manners of conveying thoughts and emotions through signing. Though the words are identical, the varied expressions communicate different messages and reflect the signer's personality. A9 provided an analogous example, likening the situation to the contrast between saying 'no' and 'noooooo' in English. A13 contributed by sharing her experience of moving across different places during her upbringing, noting that the same word can have distinct signs in different areas. The divergence in personalized signing arises due to diverse upbringing and learning environments among the signing DHH people, which can result in communication breakdowns. As A10 put it: 

\begin{quote}
    \textit{``ASL is a standardized language, and you would be using a somewhat generalized ASL, but people do have their own ways of signing and expressing themselves within ASL.''}  (A10)
\end{quote}

This benefit lies in the fact that not all human interpreters possess the level of \textit{"experience and skillsets"} required to accurately convey the nuances of these \textit{``personalized''} expressions. Our participants expressed that locating proficient human interpreters was challenging, they could be costly to hire, and their availability was not consistent. Training human interpreters demands a substantial investment of time and effort. 

\begin{quote}
\textit{`` If we can afford it, of course, a personal interpreter is better. For interpreters, ideally, to better convey the meaning of DHH individuals, they should expose themselves to more signing styles ''}(A15)
\end{quote}

It is necessary for human interpreters to not only understand individuals but also grasp the long-term context of communication (an example from A2 was that the same sign should be interpreted differently in a funeral vs. party). This becomes especially pertinent when interpreters are present for a single occasion and may lack a comprehensive grasp of the \textit{ongoing} conversation context, thus struggling to tailor their interpretation to be \textit{``personalized.''}  Three participants offered comprehensive accounts highlighting the significance of \textit{``personalized''} interpretation, especially within the K-12 educational setting where schools were responsible for interpreter expenses. They shared positive experiences tied to having a consistent interpreter over an extended period, leading to a sense of shared learning and familiarity with the interpreter's style. They mentioned the good experience of having the same interpreter for a long time who was \textit{``learning together with them'' and ``knowing them and their courses well. ''} A11 specifically mentioned that there was one interpreter she has grown up with and who knows the course well enough to provide tailored service. 

\begin{quote}
\textit{``The interpreter explained the exam questions in a way I could understand and did not tell me the right answer. I miss her so much, my favorite.''}(A11)
\end{quote}

The challenges of interpreter transitions were underscored, as adjusting to a new interpreter requires time and effort to establish synchronization. A14 added that some interpreters possess distinctive methods of signing specific words. Prior to commencing work with a new interpreter, she emphasized the necessity for in-depth communication to establish mutual understanding, thereby enhancing the efficiency of subsequent learning interactions. A more elaborate quote from her further illustrates this sentiment.

\begin{quote}
``\textit{One of my high school interpreters would always learn the course, and the terms, together with me so they would be able to explain to me very clearly, and using a way I can well understand, I was sad she eventually had health issues and had to leave... I appreciated my schools, I changed a lot of schools, and all were good enough to make sure interpreters never get changed in the middle of the semester. } '' (A14)
\end{quote}


\subsubsection{Lack of Privacy and Personal Boundaries w/ Human Interpreter} The DHH community in a certain place is often a non-mainstream minority, which includes the local interpreters who are often known by many in the community. Because most people in the community know each other, sometimes it's hard to maintain privacy. Our participants perceived the proposed technology as a way to maintain privacy when they converse with others. One participant, A13, shared an unpleasant experience she had with an interpreter which was echoed by several other participants. She mentioned that she went to Virginia to study for two years, which was not noteworthy to be discussed as it was her private matter. However, her interpreter used this information as a topic of conversation with other people, so everyone started talking about it.

\begin{quote}
\textit{``You're the interpreter. And I was like, Come on, this is my space that should have been confidential.''} (A13) 
\end{quote}

Later, A13 moved to Texas to work and study, where she found that the interpreters there sometimes also lacked control and did not uphold confidentiality regarding privacy and boundaries. This was one of the reasons why she later chose to use Video Relay Services \footnote{Video Relay Service is a form of Telecommunications Relay Service that enables individuals with a hearing loss who use American Sign Language to communicate with voice telephone users through video equipment, rather than through typed text. \url{https://www.fcc.gov/consumers/guides/video-relay-services}} instead of a face-to-face human interpreter.

\subsubsection{Lack of Access to Incidental Communication} Another benefit the participants perceived the proposed technology could provide was enabling them to have more learning and socializing opportunities \textit{``instantly.''} Many of our participants mentioned not realizing such opportunities or not being able to participate because of their lack of language access. A9 mentioned such scenarios at schools (she went to mainstream school growing up), where she felt excluded during the informal and incidental discussions where human interpreters were not provided. She perceived that such instant technology would make her feel more included and willing to speak up. Four participants found not being able to understand and react to conversations instantly \textit{``make them felt hurt,''} especially in family settings. A9 referred to a famous drawing called ``deaf dog''. \footnote{\url{http://deafcuture.blogspot.com/2015/07/7.html}} Here was the quote:

\begin{quote}
\textit{``Deaf individuals from hearing families can feel isolated without effective communication methods, leading to emotional neglect. This isolation can result in severe psychological issues... Merely expecting lipreading is cruel and ineffective, and always passively listening instead of actively participating''  } (A9)
\end{quote}

\begin{figure*}[ht]
    \centering  
    \includegraphics[width=0.9\textwidth]{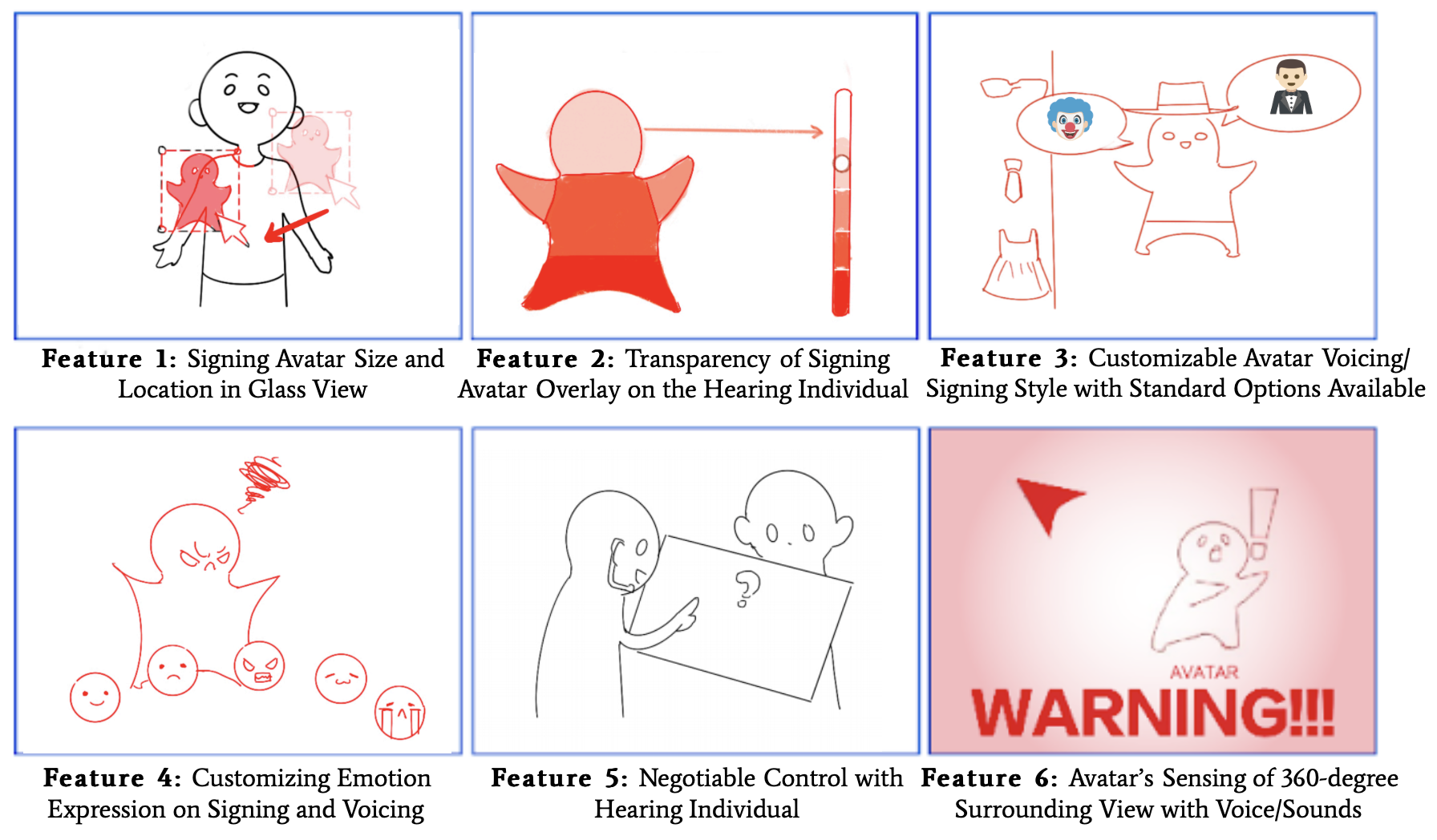} 
    \caption{Researchers Summarized Detailed Design Features suggested by DHH participants}
    \label{fig:features}
\end{figure*}


\subsection{RQ2-What overlay features are necessary to facilitate DHH-Hearing communication in MR with interpreting ability?} \label{sec:rq2}


\begin{table}[]
\resizebox{\columnwidth}{!}{%
\begin{tabular}{rcccccc}
\multicolumn{1}{l}{\textbf{}} &
  \multicolumn{6}{c}{\textbf{Participants’ Proposed Features for Customization (RQ2)} (section 4.2.1-6)} \\ \cline{2-7} 
\multicolumn{1}{l}{\begin{tabular}[bc]{@{}l@{}}\textbf{Communication} \\ \textbf{Challenges (RQ1)} \\ (section 4.1.1-4)\end{tabular}} &
  \begin{tabular}[c]{@{}c@{}}\textbf{Feature 1} \\ Signing Avatar \\ \textit{Size and} \\ \textit{Location} \\ in Glass View\end{tabular} &
  \begin{tabular}[c]{@{}c@{}}\textbf{Feature 2} \\ \textit{Transparency} \\ of Signing Avatar \\ Overlay on \\ the Hearing \\ Individual\end{tabular} &
  \begin{tabular}[c]{@{}c@{}}\textbf{Feature 3} \\ Customizable Avatar \\ \textit{Voicing/Signing} \\ \textit{Style} with \\ Standard Options \\ Available\end{tabular} &
  \begin{tabular}[c]{@{}c@{}}\textbf{Feature 4} \\ Customizing \\ \textit{Emotion} \\ \textit{Expression} on \\ Signing \\ and Voicing\end{tabular} &
  \begin{tabular}[c]{@{}c@{}}\textbf{Feature 5} \\ Negotiable Control \\ with Hearing \\ Individual\end{tabular} &
  \begin{tabular}[c]{@{}c@{}}\textbf{Feature 6} \\ Avatar’s Sensing \\ of 360-degree \\ Surrounding View \\ with Voice/Sounds\end{tabular} \\ \hline
\begin{tabular}[c]{@{}r@{}}Extra Effort Entailed in \\ Guesswork and Clarification\end{tabular} &
  x &
  x &
  x &
  x &
  \multicolumn{1}{l}{} &
  \multicolumn{1}{l}{} \\ \hline
\begin{tabular}[c]{@{}r@{}}Lack of Affordable \\ Personalized Experience \\ w/ Human Interpreter\end{tabular} &
  x &
  x &
  x &
  x &
  \multicolumn{1}{l}{} &
  \multicolumn{1}{l}{} \\ \hline
\begin{tabular}[c]{@{}r@{}}Lack of Privacy \\ and Personal Boundaries \\ w/ Human Interpreter\end{tabular} &
  \multicolumn{1}{l}{} &
  \multicolumn{1}{l}{} &
  \multicolumn{1}{l}{} &
  \multicolumn{1}{l}{} &
  x &
  \multicolumn{1}{l}{} \\ \hline
\begin{tabular}[c]{@{}r@{}}Lack of Access to \\ Incidental Communication\end{tabular} &
  \multicolumn{1}{l}{} &
  \multicolumn{1}{l}{} &
  \multicolumn{1}{l}{} &
  \multicolumn{1}{l}{} &
  \multicolumn{1}{l}{} &
  x \\ \hline
\textit{Options in Figure 1?} &
  No-Overlay &
  Partial &
  All three &
  Partial \& Complete &
  Partial \& Complete &
  All three \\ \hline
\textit{Overlay involved?} &
  - &
  Visual &
  Visual Audio &
  Visual Audio &
  - &
  - \\ \hline
\end{tabular}%
}
\caption{Summary of RQ2. Participants proposed six features that can be tailored to tackle the communication challenges identified in RQ1.}
\label{fig:summary}
\end{table}

\textit{Overview of the Three Overlay Options in Figure \ref{fig:teaser}:} From the DHH indivudual's view, the majority participants thought both `no overlay' and `partial overlay' were good options when looking at the hearing individual through the MR glasses, as drawn out by A14 in Fig \ref{fig:A14}. The `complete overlay' option was preferred by only a few for the reasons explained in the Section \ref{sec:rq3}. 

\textit{Overview of the Two Perspectives in Figure \ref{fig:teaser}:} Using the two rows in Figure \ref{fig:teaser}, DHH individuals viewed hearing individuals and how they perceived as being viewed by hearing individuals. Our participants were prompted to think from both perspectives of viewing and being viewed by the communicating pair through the MR glasses. They were able to use their knowledge of ASL in the design drawings but were at a loss when considering the voice component in the design. All participants seemed to support the use of AI-generated voice (as auditory overlays) on the DHH individual and AI-generated sign language (as visual overlays) on the hearing individual counterpart, so both parties can communicate in their preferred language - ASL and spoken English. 

\begin{table}[]
\resizebox{\columnwidth}{!}{%
\begin{tabular}{lccccc}
 & \multicolumn{4}{c}{Visual Overlay}                                                        & Audio Overlay  \\ \cline{2-6} 
 & \textbf{Face Appearance} & \textbf{Facial Movements} & \textbf{Arm/Hand} & \textbf{Torso} & \textbf{Voice} \\ \hline
\multicolumn{1}{r}{\textbf{\begin{tabular}[c]{@{}r@{}}How a DHH\\ individual sees a\\ Hearing individual\end{tabular}}} &
  Depends &
  Exaggerated for Emotion Expression &
  \begin{tabular}[c]{@{}c@{}}AI-generated \\ ASL\end{tabular} &
  \textit{\begin{tabular}[c]{@{}c@{}}No Need to \\ Overlay\end{tabular}} &
  \textit{Don’t Care} \\ \hline
\multicolumn{1}{r}{\textbf{\begin{tabular}[c]{@{}r@{}}How a DHH\\ individual is seen by\\ a Hearing individual\end{tabular}}} &
  Depends &
  \begin{tabular}[c]{@{}c@{}}Exaggerated for Emotion Expression,\\ Lips movements can be AI-generated if helpful\end{tabular} &
  \textit{\begin{tabular}[c]{@{}c@{}}No Need to \\ Overlay\end{tabular}} &
  \textit{\begin{tabular}[c]{@{}c@{}}No Need to \\ Overlay\end{tabular}} &
  \textit{\begin{tabular}[c]{@{}c@{}}Not Sure - \\ Less Access to \\ Voice Etiquette\end{tabular}} \\ \hline
\end{tabular}%
}
\caption{Overview of features applied to different body parts for `Partial Overlay.' `Partial Overlay' was preferred over `Complete Overlay' because it was perceived to facilitate communication while preserving human authenticity (to be explained in the RQ3 section). Two major modalities are involved in overlaying - visual overlay (for DHH individuals to see signs in visual communication) and audio overlay (for hearing individuals to listen to voices in auditory communication). Each option should be easily turned on and off as demonstrated with a button to cover the face in A14's figure (left drawing of Fig. \ref{fig:A14}). }
\label{fig:overview}
\end{table}

Below, we explain the common features proposed by DHH participants to address the challenges described in Section \ref{sec:rq1}. A summary of the design features is presented in Table \ref{fig:summary}. Participants discussed the required customization to fine-tune the technology for viewing the communicating pair on either side of the MR glasses. A visual illustration of customization features proposed by participants was drawn by researchers (Figure \ref{fig:features}). Using `partial overlay' as an example (as illustrated in Table \ref{fig:overview}), we found that our participants demonstrated a higher degree of expertise and comfort in designing visual overlays on hearing individuals, as opposed to their proficiency in creating auditory overlays on themselves. Their lack of confidence in this can be attributed to a lack of access to established etiquette and social norms in voice-based interactions.

\begin{figure}[!h]
    \includegraphics[width=0.9\textwidth]{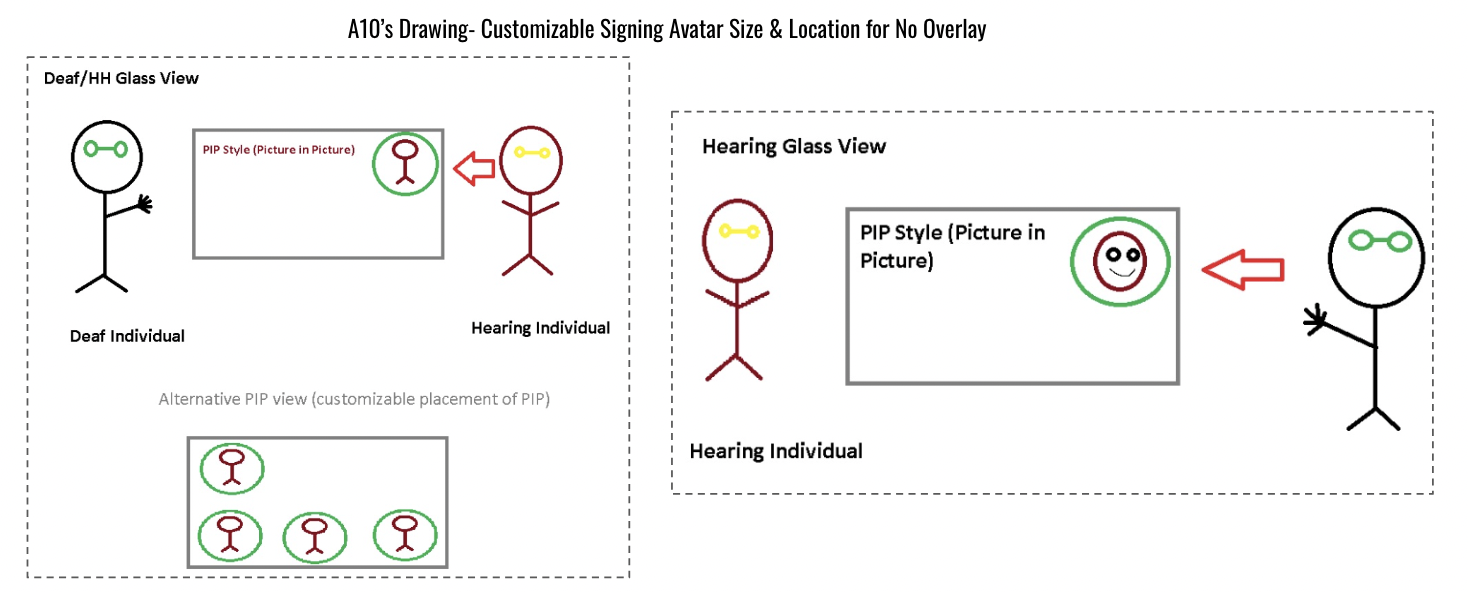}
    \caption{A10's drawing on the left shows the `no overlay' view having the adjustable location of the avatar in a `no overlay' view from a DHH individual's perspective (with signing hands and fingers). Her drawing also highlights the whole body of the interpreting avatar needs to be within view for clear visibility but not in the center, which might overlap the other individual in the glass view. From a hearing individual's perspective, the hand/arm overlay was deemed unnecessary for DHH individuals. On the other hand, the use of a face overlay, which includes AI-generated mouth movements, was considered acceptable as long as it was deemed helpful by the hearing individual.
}
    \label{fig:A10}
    \Description{}
  \end{figure}

\subsubsection{Feature 1: Signing Avatar Size and Location in Glass View} For `no overlay' specific (first column in Fig \ref{fig:teaser}), participants wanted to be able to move the avatar around in the device glass view and change the avatar size (example drawing from A10 in Figure \ref{fig:A10}). This ensures viewers to have a clear view of avatars and their signs against the backdrop with minimal eye gaze movement. It serves as a foundation for understanding interpreted messages with less effort and creates a more personalized interpreting experience. A7 emphasized the wide variety of communication needs and preferences in the DHH community, and all of these needs should be incorporated as options in the visual layout:

\begin{quote}
\textit{``For example, for one person, ASL is their first and/or only language, so they would prefer an ASL avatar/filter, whereas another person might be accustomed to English language and prefers to read the language. There is a spectrum of how everyone prefers their access to language for their own personal reasons.''(A7)}  
\end{quote}

\begin{figure}[!h]
    \includegraphics[width=.8\textwidth]{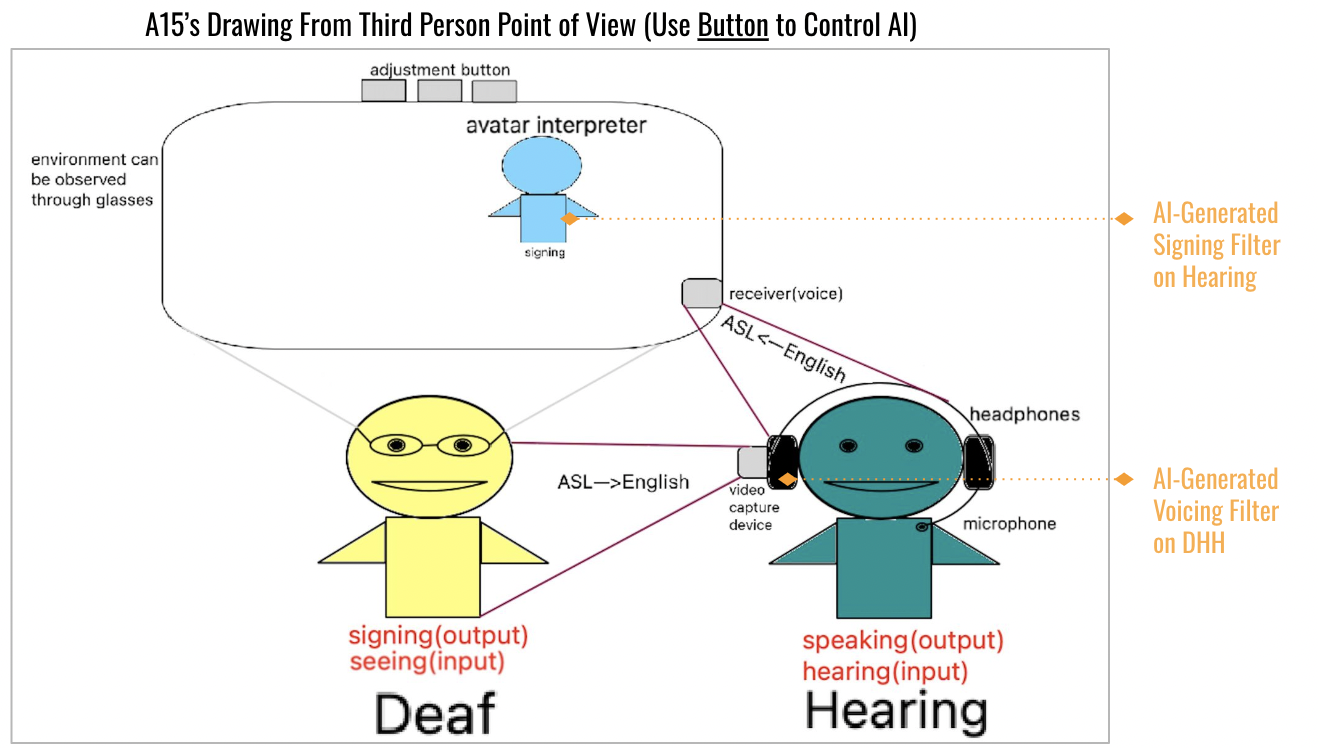}
    \caption{Participant A15's Feature Sketch showing buttons can be used to switch blue ``avatar interpreter'' visual presentation between three options in Figure \ref{fig:teaser}. Orange text added by researchers based on participants' explanations: Two main processes are involved. One is an AI-generated signing filter (interpreting voiced speech) on a hearing individual to be seen by the DHH individual. One is an AI-generated voicing filter (interpreting signs) on a DHH individual to be heard by the hearing individual.}
    \label{fig:A15}
    \Description{}
\end{figure}

\begin{figure*}[ht]
    \centering  
    \includegraphics[width=0.7\textwidth]{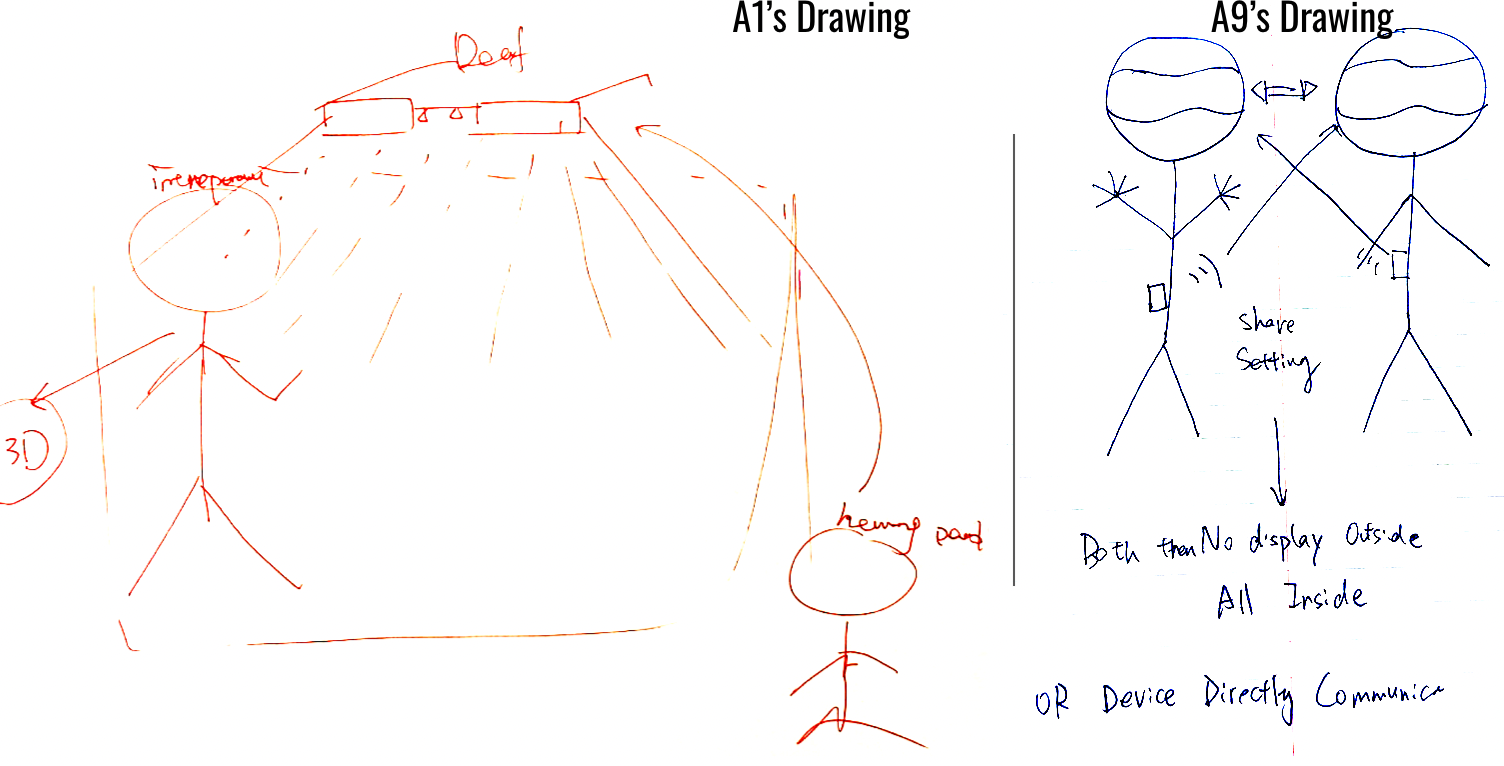} 
    \caption{Feature Sketches Showing Design Details from Third Person Point of View: The AI interpreter is a 3D avatar, both hearing individuals and DHH individuals should have control over the settings. 
    }
    \label{fig:A1A9}
\end{figure*}

A15's drawing, shown in Figure \ref{fig:A15}, emphasizes the necessity of optimal placement and visibility of the avatar and its size in the MR glasses for effective and distinctive communication from different interpersonal proximity. For location changes, the ability to move the interpreting avatar around in a 3D space was strongly preferred so as the avatar can be clearly seen, as dipicted in A1's drawing in Figure \ref{fig:A1A9}.

\begin{quote}
\textit{``It (the interpreting avatar) is located to the left or right of the screen center (depending on the viewer's preference) so that the central area can be used to observe the DHH individual's facial expression and then detect changes in the surrounding environment (using peripheral).''(A15)} 
\end{quote}

\subsubsection{Feature 2: Transparency of Signing Avatar Overlay on the Hearing Individual}  
\begin{figure}[!h]
    \includegraphics[width=\textwidth]{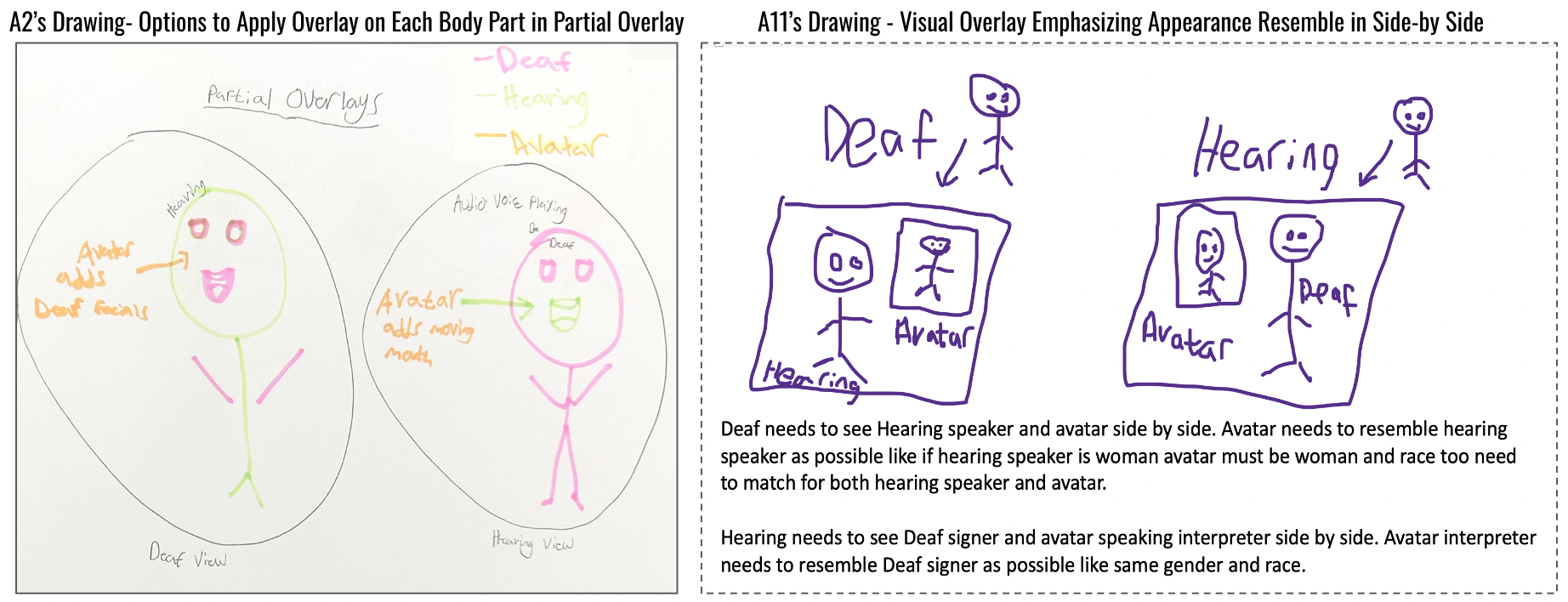}
    \caption{A2 illustrated a `Partial Overlay' from both DHH individual's and hearing individual's views. In his drawing, from DHH individual's view, hearing individual's face and torso have to be real while the facial movement and arms/hands can be AI-generated; from hearing individual's view, DHH individual's face, torso, and mouth movements must be real  while the voice can be AI-generated. A2's detailed preferences in the `partial overlay' option were shared among the majority of our participants. A depiction in A11's drawing clarified that in a `no overlay' option, the visual overlay must strive to closely resemble the original individual in order for the avatar to effectively convey the identity it represents. When the avatar does not closely resemble the original individual, it diminishes the human-human connection in the conversation and imparts a sense of third-party involvement. }
    \label{fig:A2A11}
    \Description{}
  \end{figure}

To enhance the comprehension of the ASL visual overlay generated, participants desired the ability to independently adjust the transparency of different body parts (example shown in Figure \ref{fig:A2A11}). Participants also valued the ability to cross-validate information that was AI-generated vs the original message and double-check if the AI is interpreting correctly.  This feature would also allow individuals with lipreading skills to process information in more familiar ways. However, participants preferred the torso of the hearing individual to remain authentic to maintain respectful and safe social proximity. 

As shown in Table \ref{fig:overview}, we summarized different body parts involved in the visual and audio overlay on both hearing and DHH individuals. Within various body parts (including the face and arm/hand), particular attention was placed on achieving transparency for the face. An example was from A14's drawing in Figure \ref{fig:A14} (right drawing). In her drawing, a button called ``hide face: Y \& N'' was designed to control the overlay over faces in addition to a button called ``hybrid ASL: Y \& N.'' Facial movements related to emotion are presented in \ref{emotion}. 

Regarding arms and hands, participants did not see the benefit of overlaying the signing actions of a DHH individual for the hearing audience. As exemplified in A10's illustration in Figure \ref{fig:A10}, participants believed that from a DHH individual's perspective, the entire body of the hearing individual should be overlaid with the signing avatar; whereas, from a hearing individual's viewpoint, only the face of the DHH individual should have the overlay with AI-generated mouth movements.

\subsubsection{Feature 3: Customizable Avatar Voicing/Signing Style with Standard Options Available} In addition to avatar size, location, and transparency, participants all agreed that the envisioned technology should give users options for various signing styles and facial appearance to be applied on the hearing individual, so they can understand the AI-generated signing better. They were also interested in selecting options that \textit{``matches''} the hearing individual's personality. A9 explained that for hearing individuals who are more introverted, the signing style can be slower and movements can be less animated. 

As for the facial appearance, some participants hoped that the avatar could resemble the user as closely as possible (example shown A11's drawing in Figure \ref{fig:A2A11}), while others thought it would be fun if the avatar takes on an animated form. Some DHH participants perceived that a complete face-swap could be used to maintain privacy if needed. It gives them options to hide their identity with strangers but select avatars that best represent them with friends and classmates. In a more formal context, such as a hospital, the avatar should match the individual's appearance with whom they are communicating, so \textit{``both DHH and hearing individuals don't feel a third party is involved''}.  

In addition to AI-generated signing, AI-generated voicing was exciting for most of the participants, yet some were skeptical about its value. Some believed that a standardized, unified (\textit{``neutral and professional''}) voice can reduce the chances of DHH users being mocked or discriminated against by strangers. Ultimately, the consensus among the majority was that retaining a certain level of customization is highly desirable. In brief, while standardization assures technology consistency and predictability and reduces control efforts, participants still value the ability to make choices. Viewing from the hearing individual's glasses, participants thought lip movements on a DHH individual's face could be AI-generated if the hearing individual found it helpful. As for the voice, A15 suggested asking the DHH individuals before conversing if they would choose a voice and what voice type they would prefer. She also empathized that the proposed technology must take into account that: 

\begin{quote}
\textit{``Some DHH may be uncomfortable with a (AI-generated) voice, and sometimes they are not interested in how the interpreter sounds.'' (A15)} 
\end{quote}

Another participant, A11, emphasized that the system should not mimic a DHH individual's real voice as it might sound funny and not conform to the hearing social norms, which could cause discrimination.

\begin{quote}
\textit{``I don't think that it would be a good idea to have an avatar attempt to match accent or attempt to make a sound like a deaf individual because it's a slippery slope, and people could find things funny or demeaning in the way that they interpret those accents... but we'll see'' (A11)}  
\end{quote}

\subsubsection{Feature 4: Customizing Emotion Expression on Signing and Voicing}\label{emotion} Several participants expressed a desire for increased emotional expression in the interpretation processes. Emotion filters have the potential to make translations more dynamic and captivating while also fostering trust among conversation participants and reducing misunderstandings. It is crucial that emotion filters are available on MR-enabled technology, so both conversing parties (hearing and DHH individuals) can ensure that their facial expressions and tone are appropriately \textit{``mapped''}. Some participants, particularly those with residual hearing, noted their dislike for certain AI-generated voices with monotone quality, such as Alexa, as it can hinder effective communication. They expressed a preference for AI-generated voices to convey more emotions. In Section \ref{emotionrange}, additional concerns are discussed. 

As for the AI-generated signing overlay on hearing individuals, some participants also agreed that some hearing individuals have \textit{``deaf faces''} and that the AI-generated signing filter could make facial expressions more expressive to make the conversation more engaging. A13 mentioned that some neurodivergent people might be less visually expressive than others, and the emotion filter can bridge communication for more inclusiveness. Here is one quote that explains that even with a human interpreter, some DHH individuals still pay attention to the hearing person's facial expression, indicating that emotion is necessary for understanding the conversation:  

\begin{quote}
\textit{``the interpreters' hands and whatever they sign are more like captions for me (in my current communication with hearing individuals), I still read the hearing individual's face and see what they are trying to emote and if they are experiencing breakdowns... So I personally think the 'overlay' idea was a good option for me, then I can look at the hands and face altogether, if AI can do that in the future. '' (A13)}
\end{quote}

\subsubsection{Feature 5: Negotiable Control with Hearing Individual} Our participants recognized a potential conflict between their own communication preferences and that of hearing individuals' regarding self-representation for Feature 3 and 4. Consequently, they emphasized the importance of negotiation between the parties to establish clear guidelines for categories like "never do," "probably okay," "please ask," and "go ahead." To address the issue, they suggested that both DHH and hearing individuals should set their personal preferences on their smartphones and share these settings with each other when they are in close proximity. This way, interpreting services can be automated based on both parties' preferences, as illustrated in the sample drawing from A9 in Figure \ref{fig:A1A9} (right drawing). In cases where conflicts arise, both DHH and hearing individuals can negotiate jointly by viewing a shared user interface projected in the air. A15 explained the negotiation process: 

\begin{quote}
    \textit{``There should be a well-established system ... so that both parties can come to an agreement.... Both hearing and DHH individuals should come to an agreement and feel comfortable with each other before the translation begins...In this way, both won't be confused about the change or movement of the avatar interpreter. ''(A15)}

\end{quote}

Interestingly, almost all participants found that negotiation was unnecessary for the ``no overlay'' option. This is because, in this case, the AI avatar is perceived as a service rather than a representation of a real human, which is fully controlled by the viewer. Viewers should have the freedom to move it around as they see fit. On the other hand, for the other two overlay options, particularly ``partial overlay,'' more negotiation is needed. Furthermore, there should be an option to prohibit both parties from using ``complete overlay,'' as some participants view it as \textit{``dishonest''} and believe it diminishes the human touch, as present in section \ref{humantouch}. In summary, our participants felt that most control should be given to those who are viewing for their own communication preferences.


\subsubsection{Feature 6: Avatar’s Sensing of 360-degree Surrounding View with Voice/Sounds} 
\begin{figure*}[ht]
    \centering  
    \includegraphics[width=\textwidth]{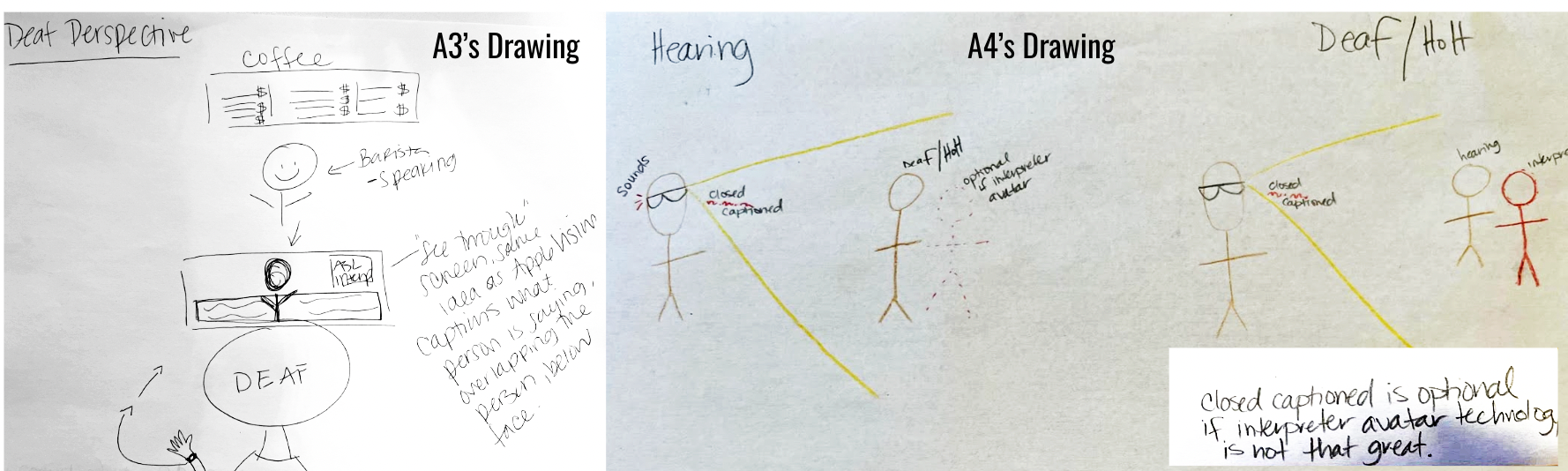} 
    \caption{Participants' Pencil Drawing on Using Auto Captioning as the communication medium (ASL to captions, and voice to captions) instead of AI-generated voicing and AI-generated signing as the ``Safer'' and ``Standardized'' Option. Hearing and DHH individuals would still be able to express themselves in their own preferred language - English and ASL. The yellow lines in A4's drawing is used to show that sound is 360-degree surrounding while vision is not.}
    \label{Fig: A3A4}
\end{figure*}

Another feature desired by our participants was an augmentation for the 360-degree surrounding view, encompassing both socializing opportunities and alerts of potential safety threats. As illustrated in Figure \ref{Fig: A3A4}, A4's drawings show the major difference between vision and auditory in developing context awareness - auditory is 360-degree surrounding, while vision is not. For example, many participants expressed a desire to be alerted to aggressive behaviors that could be concealed by the AI-generated signing hands, such as unwanted physical contact or potential harm. A9 suggested displaying notifications or warnings that can be visible to her in such circumstances, while A11 believed that the glasses should create an illusion of strangers being closer than they actually are to enhance situational awareness. Additionally, reminders for socializing opportunities with people and pets were also considered important to ensure a prompt response.

We also observed that some of our participants purposefully placed the avatar interpreter on the side instead of the center position in their design, which would enable the viewers to observe the environment for safety and socializing reasons. In their drawings, some participants purposefully reduced the size of the avatar and placed it in the top right corner. A10 provided us with an example of this design in Figure \ref{fig:A10}. This participant wanted the avatar on the screen in picture-in-picture style. According to A10, if the avatar takes up the majority of the screen, viewers using the glasses may become distracted and fail to pay attention to their surroundings, potentially leading to dangerous situations. A15 made this design decision in order to respond quickly to such incidents.

\begin{quote}
\textit{``Setting up an area on the glasses, such as the center, to observe the environment, where the avatar interpreter takes up a minimal amount of space so that the deaf individual can react quickly in case of a conflict.'' (A15)}
\end{quote}


%% file: 7-finding-2.tex
\subsection{RQ3 - What design recommendations are derived for MR-supported DHH-Hearing communication?} \label{sec:rq3} 
This section unveils design recommendations and principles for the envisioned MR technology by scrutinizing participants' concerns and questions about the technology. It further extends the detailed features explored in Section \ref{sec:rq2} and delves deeper into its real-world application, particularly within the DHH signing community - which is commented by our participants as having a broad spectrum of abilities and preferences. 

\subsubsection{Prioritizing Semantic and Emotional Authenticity in Facial and Body Movements.}\label{emotionrange} The most pivotal aspect highlighted by participants is the accuracy of semantics and emotions in translating between ASL and spoken English, and vice versa. In addition to arms/hands, facial overlay plays a crucial role in ASL overlay since facial expressions in ASL convey both language and emotions. More precisely, participants articulated the necessity for AI-generated signing hands and facial expressions overlay on hearing individuals to effectively translate their voice-conveyed emotions. Furthermore, when generating a voice overlay on DHH individuals, it should be synchronized with their signing, including the coordination of facial movements. 

Participants emphasized that emotions, neither excessive nor lacking, must be conveyed precisely as they stated,  \textit{``necessary things are enough,'' `` closely mirroring the original intent.''} To exaggerate facial movements for the hearing individuals is perceived as useful as long as it is socially acceptable in Deaf culture and as long as the resulting image does not look like a ``\textit{clown}.'' A15 specifically noted that the emotion filter should only compensate for missed emotion in the voice without altering the intended emotion: \textit{``set a limit for emotions, replicate them if they are not over the limit, and increase them if they are below.''} A3 and A2 also commented:


\begin{quote}
\textit{``I think it’s a cool idea, but every component of language must be there, the facial expressions (especially the eyebrows), body language, specific locations for the hands, lip movements, and the hands.''(A3)}
\end{quote}

\begin{quote}
    \textit{“That would be cool but you need the facial expressions too. I think I would try it. You can first remove the hearing person’s facial expressions, and then add facial expressions again for ASL. '' (A2)}
\end{quote}

Participants agreed that the proposed technology requires further research and consensus within the DHH community on how to map voice and signing parameters, acknowledging the need for additional insights and perspectives from both hearing and DHH individuals. Additionally, the technology should align the generated signing with the personality of the hearing individual; for instance, an introverted individual might exhibit fewer facial movements even when exaggerated, compared to that of an extrovert. One participant expressed a desire to use the filter to appear more "positive," while another participant mentioned using the "at work" filter for a more "professional" appearance.  Participants suggested exploring connections between signing box size, signing speed, body movement, and voice characteristics:

\begin{quote}
``\textit{Some people have larger signing boxes while some have smaller ones. Maybe when someone speaks louder, the machine-generated signing box can be bigger, and when the person speaks quieter the movements can be smaller. The signing speed and body movement probably can present something else related to voice. But again, I don't know a lot about voices and that needs more insights from both hearing and DHH individuals.''(A14)}
\end{quote}

\subsubsection{Enhancing Interaction Authenticity via Eye Contact and Unfiltered Face.}\label{humantouch} In our study, another significant aspect raised by our participants is the importance of a \textit{``human touch'' in technology-mediated conversations}. The proposed designs must convey a sense of "seeing the person" and engaging with genuine humans while avoiding any impression of concealment. Some participants linked this "human touch" to authentic human connections and virtues. They emphasized that technology should not hinder, but rather enhance real human interactions and associated merits.

Regarding the three options, all participants found a ``partial overlay'' acceptable (with some even preferring it as their top choice). In the case of a ``partial overlay,'' all participants noted that, from the perspective of the DHH individuals' view, at least the eyes should remain unfiltered and they should have access to the unfiltered physical self of the other party, especially the unfiltered face. This approach allows DHH individuals to maintain eye contact and make sure they are looking at a \textit{real person}, as opposed to constantly shifting attention like in ``no overlay.'' In addition, a few participants expressed the belief that employing a full coverage ASL overlay for hearing individuals (to simulate signing) or vice versa is \textit{``dishonest''} and should be avoided, as opposed to ``partial overlay.''  \textit{``I don’t want someone to be a fake deaf person. They must be real, period.''} (A3)

\subsubsection{AI-generated Voice/Signing Customization.} Nearly all participants expressed enthusiasm about the options to customize AI-generated voice/signing, although their opinions regarding the degree of customization differed. Some participants believed that there should be a high degree of customization, allowing them to adjust the technology to best suit their needs, while others thought that maintaining a certain level of consistent standards can help more users get started more comfortably without the need to ``\textit{depend on others on the picking a voice.}''  Some participants indicated that they would seek input from friends and family to assess whether the generated voice suits them, which is helpful, but at the same time, an independent decision is traded in. 

Supporting examples highlighted the need for a more transparent and understandable AI-generated voice. Many initially excited participants found themselves at a loss when asked to describe preferred voice attributes. A13, a transgender masculine participant, expressed a desire for a more "masculine" voice option, as opposed to the typical "feminine" voice that people often assume based on outward appearance. 
Another participant shared, 

\begin{quote}
    \textit{"The other day, my profound deaf friend asked me to listen to her car to see if it sounds weird and if the car is having a problem... that's the kind of sound I can discern, but my friend cannot hear it... You know what I mean, hearing less means less control... I want to be able to understand it then select what I want, always good to have options ."} (A15)
\end{quote}

During our study, many participants discussed how their limited access to voice and voice-based interactions hindered their ability to select their own AI-generated voice. Describe voice characteristics, and provide input on design features involving voice input and output. Granting control over their AI-generated voices is vital for DHH individuals as it influences how they are perceived by others. Participants expressed curiosity about the social appropriateness of AI-generated voices. A4 admitted uncertainty about hearing world etiquette, which affected their design input. A15 suggested conducting a survey to identify commonly preferred voice features among the hearing population. This insight could help explain AI-generated voices and aid DHH individuals in their decision-making on which voice to pick. 

Two participants mentioned that interpersonal proximity in conversation is different between two hearing people using voice and between two DHH individuals using signing. Therefore, it is necessary to have the hearing and DHH conversing parties agree on a set of preferences comfortable to both before the conversation exchange begins.

\begin{quote}
    \textit{``As for the interpersonal proximity, this should be set before the interpretation begins so that both parties are at a comfortable and suitable space.'' (A15)}
\end{quote}

Some participants, out of their desire for a "safe" and "socially approved" outcome, preferred \textit{standard voicing/signing generation} over \textit{personalized voicing/signing generation}. For example, using "Alexa" for everyone's AI-generated voice alleviates DHH individual's burden of voice control, which is challenging given their limited access to and understanding of sound. Another example from A3 and A4, illustrated in Figure \ref{Fig: A3A4}, show that they preferred captions instead of voice to avoid the lack of control over AI-generated voice. Auto-captioning serves as the intermediary for communication, facilitating the conversion of ASL to captions and spoken language to captions. This allows both hearing and DHH individuals to express themselves in their preferred languages, be it English or ASL, and receive the other party's message in captions.

\subsubsection{Adjusting Technology to Accommodate a Wide Spectrum of Voicing \& Hearing Preferences} An important feature that every participant mentioned is the adaptability and adjustability of the technology tools that can be tailored to individual's speaking and hearing preferences. They mentioned that being DHH does not mean not being able to voice, and being able to voice does not mean good hearing ability. Our participants mention they (and their friends) have varied voicing abilities. Some stated they can voice (because of the long therapy they have been going through growing up) even if they barely hear at all. 

Besides the ability to voice, some are comfortable using their voice and some are not; especially when auto-captioning services are turned on, they feel uncomfortable in voicing and are impatient when their voice gets recognized incorrectly by technologies. As for their hearing ability (with hearing aids), some mentioned they can hear certain frequencies of sound or lip-read to consume the information being conversed, which made them inclined to the option to listen to speaking individuals and not to overlay the speaking person's mouth movements. 

The majority of our participants mentioned they would like to use ``Apply Emotion Filter'' and ``Generate Personal Signing/Voicing Style'' depending on whom they are communicating with.  Overall, they were more willing to use these two features with family/friends but not at workspace/school. They had concerns and doubts about how well AI can do to represent them perfectly. One reason is their concerns about the social appropriateness of generated voice as discussed before. The second is that they perceived applying emotion filters to their faces/voices to potentially change the message being conveyed. For workspace/school, it is more crucial to have the \textit{``exact information''} transferred and be interpreted the same by different people. In comparison, communication with family/friends allows emotion as well as \textit{information} to flow through conversation and be playful.

In summary, our participants' responses shed light on the variations in DHH individuals' ability to hear and voice as well as their willingness to perform such functions for themselves. They suggest our proposed design to include a wide range of options in addition to a ``standardized'' setting (e.g. everyone's AI-generated voice in Alexa) which can make DHH individuals feel ``safer'' and inclined to rely on the technology. The ``partial overlay'' option was also perceived as more adaptable to a variety of communication preferences among the community. The envisioned technology needs to be inclusive among the DHH community to make people feel comfortable in using it as a ``personal gadget'' and thereby, accept/adopt the technology.

\begin{quote}
    \textit{``The key word is experimentation, try different layout and filter. There are many different individual preferences. There’s a lot of variation and many possibilities.'' (A4)}
\end{quote}

\begin{quote}
    \textit{``For example, for one person, ASL is their first and/or only language, so they would prefer an ASL avatar/filter, whereas another person might be accustomed to English language and prefers to read the language. There is a spectrum of how everyone prefers their access to language for their own personal reasons.'' (A7)}
\end{quote}

\subsubsection{Respecting ASL Norms While Adhering to Hearing Social Norms} Participants acknowledged the technology's potential utility but stressed that interpretation transcends literal word-for-word translation. They emphasized the importance of recognizing that ASL is a distinct language with unique characteristics compared to spoken English. The technology should strive to make the generated voice sound natural within the norms of the hearing community while also respecting the ASL expression norms and habits of DHH signing individuals. It's crucial for people, DHH or hearing, to be mindful of the these considerations when using the technology.  
\begin{quote}
    \textit{``ASL and English are different languages, so there is no absolute fast or slow in expressing oneself or things. It can be perceived that ASL is faster when describing images or actions, but English is faster when describing technical terms or logical events. Our design should slow down sign language when describing actions and English when describing logic, so that some kind of balance can be achieved. ''(A15)}
\end{quote}


Participants highlighted that DHH-Hearing interaction mediated by human interpreters can increase awareness among hearing individuals about the identity and experiences of the DHH community. This interaction can also lead to some hearing individuals to learn sign language. However, there are concerns that this identity representation and awareness education might be lost if a ``complete overlay'' is used. Participants emphasized the importance of both parties having the opportunity to adapt to and collaborate with the technology, similar to their interaction with human interpreters. They stressed that it is essential for participants to understand how to effectively engage with the technology and comprehend its functionalities and limitations. This understanding is seen as a way to promote awareness of accessibility in general and ASL, Deaf culture in particular. More research on hearing and DHH individuals collaborating with interpreter teams can be found in \cite{rui2022online}.

\begin{quote}
    \textit{``DHH individuals need to understand how the avatar-interpreter functions and its connection with hearing individuals. So do hearing individuals. So that both can make the interpretation work properly.'' (A14)}
\end{quote}

%% file: 8-discussion.tex
\section{Discussions}

Our findings underscore the pressing need for advancements in \textit{interpretation} as a design space and material within the field of HCI. Interpretation goes beyond word-for-word translation to include multilingual interfaces, multimodal interactions, and intracultural nuances \cite{meulder2021interpreting}. Our findings show effective interpretation technology must convey emotions, personalities, and regional dialects, and adapt to learners' knowledge and comfort levels. It should also offer diverse signing and voicing styles for clarity in communication.

Our participants' envisioned technology is proposed to address major issues with human interpreters, such as prescheduling, unqualified service, boundary, and privacy issues, which can be partially resolved by introducing AI with interpreting abilities and embedding them in MR settings. This new technology can also offer features that human interpreters might find challenging, such as representing participants in their preferred way in group settings and personalizing the interpreting experience more effectively. 

These features and design recommendations are not meant to replace human interpreters but to be supported by them. 
AI involvement in interpreting can vary from completely human-generated with no oversight to AI-generated with no oversight. In between, one can have a spectrum from human-generated with AI tools to AI-generated with real-time or offline human edits. At present, AI is good at the information level but is less capable at the communication level that involves shared cultural and linguistic understanding, which is not well-represented at the AI level. How human interpreters' practices will change and face new challenges with new technology remains an open area for future exploration. For example, Ang et al.'s work found that video conferencing systems like Zoom are less accessible for human interpreters and require strategies to work effectively \cite{rui2022online}.

\subsection{Design Implications - Embodied Authenticity of Customized Communication} 

Pedersen, who first mentioned "embodied authenticity," argues that AR interfaces must accommodate humans physically and mentally. Achieving an immersed "self" requires multisensory interaction that mirrors real-world communication and learning \cite{duin2019human, pedersen2009radiating}. Our study found that "embodied authenticity" can be achieved through customized communication, such as AI-generated voicing for DHH individuals, signing for the hearing individuals, and emotional expression intensity for both groups. Accurate mapping between modalities and movements is crucial for authenticity. Below, we provide design implications for supporting "embodied authenticity" in communications, which could be insightful beyond mixed-hearing ability parties. 



\subsubsection{Customizing Body Movements}  
\textit{(1) Controlling Partial Overlays via Different Modalities and Body Movements}. Among the overlay options, `partial overlay' serves as an example of fine-grained self-presentation control needed, encompassing multiple body parts and modalities (see Figure \ref{fig:overview} and Figure \ref{fig:A2A11}(A2)). Future research could explore modalities that facilitate user identity representation and control, such as VR handwriting for login and identity input \cite{lu2020fmkit}. Our findings also suggest the need for giving users the option to adjust the locations of interpreting avatars in a 3D space to enhance respectful and comfortable communication in scenarios without overlays.

\textit{(2) Conveying Personality Representation}. As participants noted, different physical movements can represent varying personalities and inner selves (e.g., extroverted individuals using larger movements and signing with more variation in speed). Additional speed and timing parameters of ASL movements including sign duration, transition time, differential signing speed, pause length and frequency can be explored \cite{al2021different}. Further research is needed to investigate how personality can be conveyed across different modalities and its relationship with space and timing, such as AI-generated voices displaying diverse emotions and traits \cite{zhang2021social}. Given the fluidity of user identities and self-representation needs, and avatars aren't bound to reality and can reflect intersectional and changing identities \cite{mack2023towards}, we recommend systems include features enabling the portrayal of diverse personalities through various modalities and movements.

\textit{(3) Negotiating the Communication Settings}. Our design focuses on two-way communication (two rows in Figure \ref{fig:teaser}). As demonstrated in the RQ2 discussion, negotiations between the two conversing parties may be necessary to solve potential conflicts in preferences before the actual interpretation begins. Our research (Feature 5) advocates for negotiable AI control to address conflicts related to self-representation needs, building on prior work concerning privacy and social boundaries in AI-mediated human interactions (e.g., \cite{zheng2022ux}). Users should have options to set and adjust their willingness to be filtered by others for each aspect of communication. Different negotiable levels, such as "No Overlay Allowed," "By Request," and "Go Ahead," should be offered to facilitate self-representation alignment for both parties. 

\subsubsection{Customizing Voice}
For most DHH participants, a common challenge lies in the creation of audio, such as AI-generated voices, that accurately represent theirs' while perceived as acceptable for those who can hear. This challenge arises due to many DHH people have limited or no exposure to voice and are not familiar with social norms centered around voice-based communication. While our research predominantly centers on Deaf as a minority culture, we recognize the importance of leveraging these findings to promote awareness among hearing people so they can learn to represent themselves appropriately when communicating with DHH people. 


\textit{(1) Explaining AI Generation Outcome from Receiver Perceptive}. Our participants recommended the use of a survey to identify preferred voice features for DHH people to enhance AI-generated voice selection for the DHH community. This highlights the need to clarify the modality (e.g., voice pitch) and its perception by the audience (e.g., human-preferred pitch range and emotional impact). We build upon \cite{zhang2022s} that suggest avatar customization outcome should be accessible to blind users using alt text. Additionally, as illustrated in the participants' drawings and RQ1, visual range is limited by an individual's field of view, while the auditory range can extend beyond the user's direct line of sight. Such valuable spatial information need to be explained in the generation. Throughout the process, technology should increase audience awareness of `collective access' \cite{mcdonnell2023easier,mack2021mixed} and genuine intercultural understandings to minimize potential misunderstanding and discomfort \cite{chen2023my, bala2023towards, DIS23}.


\textit{(2) Empowering Presenter's Control through Explanation}. Our DHH participants wanted to involve their hearing friends in choosing matching generated voices and selecting signing styles for hearing individuals. However, this reliance on human input raises dependency concerns. Also, prior research has shown extra effort in managing self-representation required by users with disabilities in videoconferencing \cite{tang2021understanding}. Therefore, future technology should reduce user effort in self-representation and promote decision independence. It underscores the role of machines in comprehending and articulating both within and about diverse cultures.



\subsection{Researchers' Reflections on Accessible Study Design} 

Making human-centered methods accessible requires careful consideration and iteration, as Mack et al. noted \cite{mack2022anticipate}. Below, we reflect on our lessons learned on making study method (in)accessibile. 

\subsubsection{Zoom vs In-person Participation} In our study, some participants joined remotely while others attended in person. Remote participation allowed for greater time flexibility and geographical diversity but also increased access barriers related to communication technology. Zoom sessions were longer, averaging 20 minutes more than in-person sessions, with two participants' studies totaling four hours. The primary reason was that interacting with Zoom screens caused more fatigue, required breaks, and needed additional attention for effective communication, including using features like pinning videos, resizing thumbnails, and adjusting lighting. Additionally, the drawing materials used in session 2 differed between in-person and Zoom participants. Participants were asked to prepare their preferred drawing materials in advance, which they appreciated. This reminds future researchers to consider how communication technology diversifies participation and to plan accommodations for materials, space, and time. 

\subsubsection{Culturally Aligned Visual Communication in Design Activities} 
Another major takeaway from our study is that to make technology designs accessible for DHH individuals, the explanation of design ideas and participation in design activities should be visual. The primary outcome of our practice session, illustrated in Figure \ref{fig:teaser}, effectively explains MR concepts visually, making them more understandable for our population. Participants commented that this visual approach was more effective than fingerspelling, which often doesn't convey much meaning. Most participants' drawings were clear and easily understood by other likely due to their mode of communication—sign language is a visual language and Deaf culture emphasizes the importance of visual communication. Each participant's ideas were clearly illustrated through drawings and could be reviewed effectively.  Additionally, DHH individuals express themselves well visually, which makes design review possible. For other user populations that are less visual, whether session 3 will work remains to be further explored. Therefore, future research should consider ways for participants to understand and participate that align with their cultures and modalities.

\subsection{Limitations}
Our study has several limitations that must be acknowledged. First, the limited sample size and the relatively young average age of participants, primarily millennials and Generation Z individuals may have influenced the results. Their familiarity with avatars and filters through social media could have made them more comfortable with the concept, potentially skewing their perceptions and responses to the ASL avatars presented in the study. Furthermore,   The second limitation is related to reviewing other participants' design ideas in the last session. The time spent on this activity and the number of designs reviewed was not consistent as some participants were faster than others in completing their tasks in previous sessions. The uneven number of designs reviewed might impact the results which needs to be further studied in the future. The third limitation is that our technology demonstration chosen in Session 1 may have influenced the findings presented in our study. Future research may explore to what extent our results may be impacted by changing technology stimuli.

Another limitation is that our research primarily focused on the perspective of the DHH signing community and extrapolated findings to discuss insights for those who can hear. However, this may not reflect the genuine reality as many DHH individuals have grown up navigating both hearing and Deaf cultures, as evident in some of our quotes. DHH individuals who do not identify as culturally Deaf or older adults who have a loss of hearing later in life or deaf people who do not sign may prefer captioning tools instead of signing avatars. Consequently, the findings may only partially represent the perspectives of DHH individuals who do not sign or those less exposed to avatars and filters. In contrast, hearing individuals may lack the same level of experience in learning and adapting to another culture. As highlighted in recent research \cite{mcdonnell2023easier}, DHH-hearing established groups have developed social accessibility norms over time that align with their specific relational contexts, emphasizing the need for a gradual process. Our findings also emphasize the importance of users' active engagement with the technology. Moving forward, a crucial next step is to explore the design aspects involving hearing individuals with or without experiences interacting with users with disabilities. This presents an opportunity for intra-cultural dialogue to enhance accessibility literacy. Moreover, we propose using \textit{Drama} to ``allow participants to enact the roles of different stakeholders to understand the relation complexity involved'' and \textit{Deliberation} for ``collective problem-defining, problem-solving, and decision-making, building on the participants’ shared experiences and reflection'' \cite{pschetz2019autonomous}. 

%% file: 9-conclusion_and_limitation.tex
\section{Conclusions} The findings from this study contribute to ongoing efforts to create more inclusive communication and learning technologies. Using participatory approaches, we explored the innovative design of MR (Mixed-Reality) empowered by AI that can interpret to enhance face-to-face communication between DHH and hearing people, featuring a signing avatar for hearing individuals and a voicing avatar for DHH individuals. Throughout the study, DHH participants expressed enthusiasm about the envisioned technology to improve communication with hearing individuals in various aspects of their lives, including in work and family settings. The concept of instant, personalized, and emotion-rich communication resonated positively with the participants, who recognized its potential to make DHH-Hearing face-to-face communication more seamless. While acknowledging the benefits, DHH participants also expressed concerns about how such AI-generated voices might behave and adapt to auditory social norms. 
Exploring and evaluating more fine-grained controls for customizing the AI-generated avatar could be a focus of future research. We discuss design recommendations for future MR-supported communication and learning for individuals with mixed abilities, extending beyond just the hearing and DHH communities.